\documentstyle[pre,aps,epsf]{revtex}
\begin{document}
\draft
\author{C. Bagnuls\thanks{%
e-mail: bagnuls@spec.saclay.cea.fr}} 
\address{Service de Physique de l'Etat Condens\'{e}, C. E. Saclay, F91191 Gif-sur-Yvette Cedex, France}

\author{ 
C. Bervillier\thanks{%
e-mail: bervil@spht.saclay.cea.fr} }
\address{Service de Physique Th\'{e}orique, C. E. Saclay, F91191 Gif-sur-Yvette Cedex, France}
\date{January 11, 2002}

\title{Classical-to-critical crossovers from field theory}
\maketitle

\begin{abstract}
We extent the previous determinations of nonasymptotic critical behavior of Phys. Rev
{\bf B32},  7209 (1985) and {\bf B35}, 3585 (1987) to
 accurate expressions of the complete classical-to-critical
crossover (in the 3-d field theory) in terms of the temperature-like scaling
field (i.e., along the critical isochore) for : 1) the correlation length,
the susceptibility and the specific heat in the homogeneous phase for the $n$%
-vector model ($n=1$ to $3$) and 2) for the spontaneous magnetization
(coexistence curve), the susceptibility and the specific heat in the
inhomogeneous phase for the Ising model ($n=1$). The present calculations
include the seventh loop order of Murray and Nickel (1991) and
closely account for the up-to-date estimates of universal asymptotic
critical quantities (exponents and amplitude combinations) provided by Guida
and Zinn-Justin [J. Phys. {\bf A31}, 8103 (1998)].
\end{abstract}

\section{Introduction}
The asymptotic critical behavior, characterized by universal quantities
(exponents and amplitude combinations), is now theoretically well
established \cite{4948,3533} with accuracy \cite{4211}. However, the
comparison of the theoretical results with experimental or numerical data
is made easier when the theoretical expressions are extended into regimes
where the asymptotic pure scaling breaks down \cite{2720,732,733,736}
(calculations done far away from the critical point, characterized by
nonuniversalities and including eventually crossover phenomena, see reviews \cite{4948,178,3534,4748,5046}). This
extension has appeared necessary notably when measurements on colloids \cite
{3053}, but also on complex systems such as ionic fluids \cite
{3096} or polymers \cite{5186}, seemed to yield strong nonuniversalities in
approaching the critical point. Indeed theoretical studies have suggested
that, in some cases, those nonuniversalities could be due to a phenomenon of
``retarded criticality'' which characterizes measurements done outside the
asymptotic critical domain \cite{177,4627}. Several recent
theoretical (and/or numerical) studies have as well explicitly considered
the evolution of effective exponents with emphasis on their monotonic or
non-monotonic character \cite{5046,3871,4073,4215,4216,5052}. Furthermore
the description of the classical-to-critical crossover for Ising systems is
not yet clear-cut \cite{4732,3504}. For these reasons and because
our preceding determination of nonasymptotic critical behavior from field
theory \cite{733,736} did not yield continuous functions covering an entire
crossover region, it seems to us useful to consider again those calculations
in order to (see also \cite{4073,4323,foot18}):

\begin{enumerate}
\item  extend them to a complete account of the classical-to-critical
crossover which characterizes the framework of field theory \cite
{440,4595},

\item  include the seventh order series for the critical exponents
determined by Murray and Nickel \cite{312} in order to account as closely as
possible for the up-to-date estimates of universal asymptotic critical
quantities (exponents and amplitude combinations) provided by Guida and
Zinn-Justin \cite{4211} (referred to in the following as GZ).
\end{enumerate}

In the previous work \cite{733,736}, and contrary to an initial attempt \cite
{732} regarding the homogeneous phase ($n=1$), we provided only continuous
expressions of $t$ valid for $t\lesssim 10^{-2}$ ($t$ is the
temperature-like scaling field which is proportional to the absolute value
of the reduced critical temperature $\left| T-T_{c}\right| /T_{c}$). The
crossover was not completely described because it was thought at that time
that the field theoretical framework had a range of validity strictly
limited to the first correction to scaling term. Consequently, the practical
limit of physical validity of the functions was imposed by the range of $t$
where the second correction to scaling term specific to field theory becomes
non negligible and this occurs \cite{733,736} about $t\simeq 10^{-2}$. Since
then, it has appeared that the range of validity of field theory could be
much larger than that and even could cover the entire classical-to-critical
crossover region \cite{4627} that it describes. Somehow it is interesting to
give expressions valid in the entire crossover region if only because it may
be compared to other kinds of classical-to-critical crossovers either
experimental \cite{4748} or from numerical studies \cite
{5046,4216,4732,3869,4231} which, under some particular conditions,
 are identical \cite
{4073,4215,4323,3869,4355} to the field theoretical form 
(but see also our comment in \cite{3504}).

For technical reasons, in the preceding work \cite{733,736} we did not
constrain our theoretical expressions to include very closely the estimates
of universal asymptotic critical quantities of that time (with their error
bars) so that uncertainties were underestimated and then our estimates of
the correction amplitude ratios were, presumably, also not firmly
determined. Moreover small errors existed in the preceding study of the
inhomogeneous phase \cite{736} (as indicated elsewhere \cite{4073,2376,2376bis}%
) which have been eliminated from the present work (nevertheless, we have explicitly
verified (see fig. 1 of ref. \cite{2376bis}) that the errors have had no important consequence on the final
results as it could be clearly deduced from a comparison of our
estimates of universal amplitude-combinations \cite{736} with those of Guida
and Zinn-Justin \cite{3639,4211}).

\section{Principle of the calculations}

\subsection{Brief reminder\label{remind}}

As in the preceding work \cite{733,736}, and using the same resummation
method, we have considered the correlation length $\xi \left( t\right) $ (in
the homogeneous phase $T>T_{c}$ for the $n$-vector model with $n=$1 to 3),
the susceptibility $\chi \left( t\right) $ and the specific heat $C\left(
t\right) $ (in the homogeneous phase $T>T_{c}$ with $n=$1 to 3 and in the
inhomogeneous phase $T<T_{c}$ with $n=$1) and the coexistence curve
(spontaneous magnetization) $M\left( t\right) $ (in the inhomogeneous phase
with $n=$1).

For practical reasons, it is useful to fix our notations relative to the
actual asymptotic critical behaviors (i.e., in terms of the physical
variable $\tau =\frac{T-T_{c}}{T_{c}}\rightarrow 0^\pm$ instead of $t\rightarrow
0$, see also section \ref{USEOF}):
\begin{mathletters}
\label{eq:notat}
\begin{eqnarray}
\xi \left( \tau \right) &=&\xi _{0}^{\pm }\;|\tau |^{-\nu }\left[ 1+a_{\xi
}^{\pm }\;|\tau |^{\Delta }+O\left( |\tau |^{2\Delta }\right) \right]
\label{eq:notata} \\
\chi \left( \tau \right) &=&\Gamma ^{\pm }\;|\tau |^{-\gamma }\left[
1+a_{\chi }^{\pm }\;|\tau |^{\Delta }+O\left( |\tau |^{2\Delta }\right)
\right]  \label{eq:notatb} \\
C\left( \tau \right) &=&\frac{A^{\pm }}{\alpha }\;|\tau |^{-\alpha }\left[
1+\alpha a_{C}^{\pm }\;|\tau |^{\Delta }+O\left( |\tau |^{2\Delta }\right)
\right] +B_{cr}  \label{eq:notatc} \\
M\left( \tau \right) &=&B\;|\tau |^{\beta }\left[ 1+a_{M}\;\left| \tau
\right| ^{\Delta }+O\left( |\tau |^{2\Delta }\right) \right]  \label{eq:notatd}
\end{eqnarray}
\end{mathletters}
in which $\alpha $, $\beta $, $\gamma $ and $\nu $ are the critical
exponents, $\Delta $ (also denoted by $\theta =\omega \nu $ by GZ)
is the correction exponent, $\xi _{0}^{\pm }$, $\Gamma ^{\pm }$, $A^{\pm }$
and $B$ are the leading critical amplitudes and $a_{\xi }^{\pm }$, $a_{\chi
}^{\pm }$, $a_{C}^{\pm }$ and $a_{M}$ are the (confluent) first-correction
amplitudes, finally $B_{cr}$ is a critical background. One usually restricts
the consideration of the critical singularities to small values of $t\propto
|\tau |$ as it is implicitly assumed in Eqs.\  (\ref{eq:notat}). The
obtention of nonasymptotic critical behavior supposes the explicit
consideration of non-necessarily small values of $t$

Let us suppose that we want to calculate the susceptibility $\chi $ as
function of the (non-necessarily small) temperature-like scaling field $t$.
Calculations of such a nonasymptotic critical behavior from (the massive)
field theory (in three dimensions) \cite{732,733,736} present the following
features (additional details may also be found elsewhere \cite{4952}):

\begin{enumerate}
\item  the function $\chi \left( t\right) $ is primarily performed under the
implicit form because the quantities $\chi $ and $t$ are primarily given as
perturbation series in powers of the renormalized coupling parameter $g$ (up
to fifth \cite{80,733} or sixth \cite{736} order): the functions $\chi
\left( g\right) $ and $t\left( g\right) $ are resummed for $g$ varying in
the range $\left] 0,g^{*}\right[ $ where $g^{*}$ is the zero of the Wilson
function (or the ``$\beta $-function'') $W(g)$ also primarily given as power
series of $g$ (up to sixth order \cite{323}; $g^{*}$ is the fixed point
value of $g$).

\item  \label{Fit}The consideration of discontinuous values of $g$ is a
compelling need of the numerical resummation procedure. Consequently fitting
an ad hoc function of $t$ to the calculated points eliminates the auxiliary
variable $g$ and provides us with the final expression of $\chi \left(
t\right) $ as the explicit continuous function of $t$ we are looking for in
the range $\left| t\right| \in \left] 0,+\infty \right[ $.

\item  The actual calculation of the quantities of interest [like $\chi
\left( g\right) $] for values of $g$ close to $g^{*}$, at which point they
are singular (due to the critical singularity we expect to closely
reproduce), requires expressing them under an integral representation like: 
\begin{equation}
\chi \left( g\right) =\chi \left( y_{0}\right) \exp \left[
-\int_{y_{0}}^{g}\;dx\;\frac{\gamma \left( x\right) }{v\left( x\right)
W\left( x\right) }\right]   \label{eq:integrale}
\end{equation}
in which $\gamma \left( g\right) $ and $v\left( g\right) $ are not singular
at $g^{*}$ and are primarily given as power series of $g$ (up to seventh
order \cite{312}). Especially, $\gamma \left( g^{*}\right) $ and $v\left(
g^{*}\right) $ provide the field theoretical estimates of the critical
exponents $\gamma $ and $\nu $. Only the elementary series $\gamma \left(
g\right) $, $v\left( g\right) $ and $W\left( g\right) $ are resummed using
the sophisticated method mentioned in the following step [the value $y_{0}$ is chosen
small enough to allow a direct simple summation of the series $\chi \left(
y_{0}\right) $].

\item  \label{Leroy}To sum perturbation series like $\gamma \left( g\right) $%
, $v\left( g\right) $ or $W\left( g\right) $ for a given value of $g$ a
Borel--Leroy transformation is used, combined with a conformal mapping. An
estimation of the error is deduced from the observation of the convergence
properties of the series when varying the free parameters of the
transformation. This leads us to fixe those parameters (resummation
criteria) in such a way as to obtain a combination of the error bounds on
e.g., $\gamma \left( g\right) $, $v\left( g\right) $ and $W\left( g\right) $
which gives a kind of envelope for $\chi \left( g\right) $ via two functions 
$\chi _{\max }\left( g\right) $ and $\chi _{\min }\left( g\right) $ (and
similarly for $t\left( g\right) $).

\item  Since the critical singularities are similar in the two phases of the
transition, the calculations in the inhomogeneous phase ($T<T_{c}$) do not
require the consideration of new series for the exponents compared to the
homogeneous phase ($T>T_{c}$). Hence the same three series $\gamma \left(
g\right) $, $v\left( g\right) $ and $W\left( g\right) $ express the critical
singularities via integrals similar to that given in (\ref{eq:integrale}),
only new critical amplitude functions of $g$ (hence not singular at $g^{*}$)
must be calculated \cite{80,736} and summed using the transformation
mentioned in step \ref{Leroy}.
\end{enumerate}

\subsection{Improvements to the previous work and presentation of the results}

\subsubsection{The fitting procedure}

\begin{figure}
   \centerline{ \epsfxsize=12in \epsfbox[-150 200 750 700]{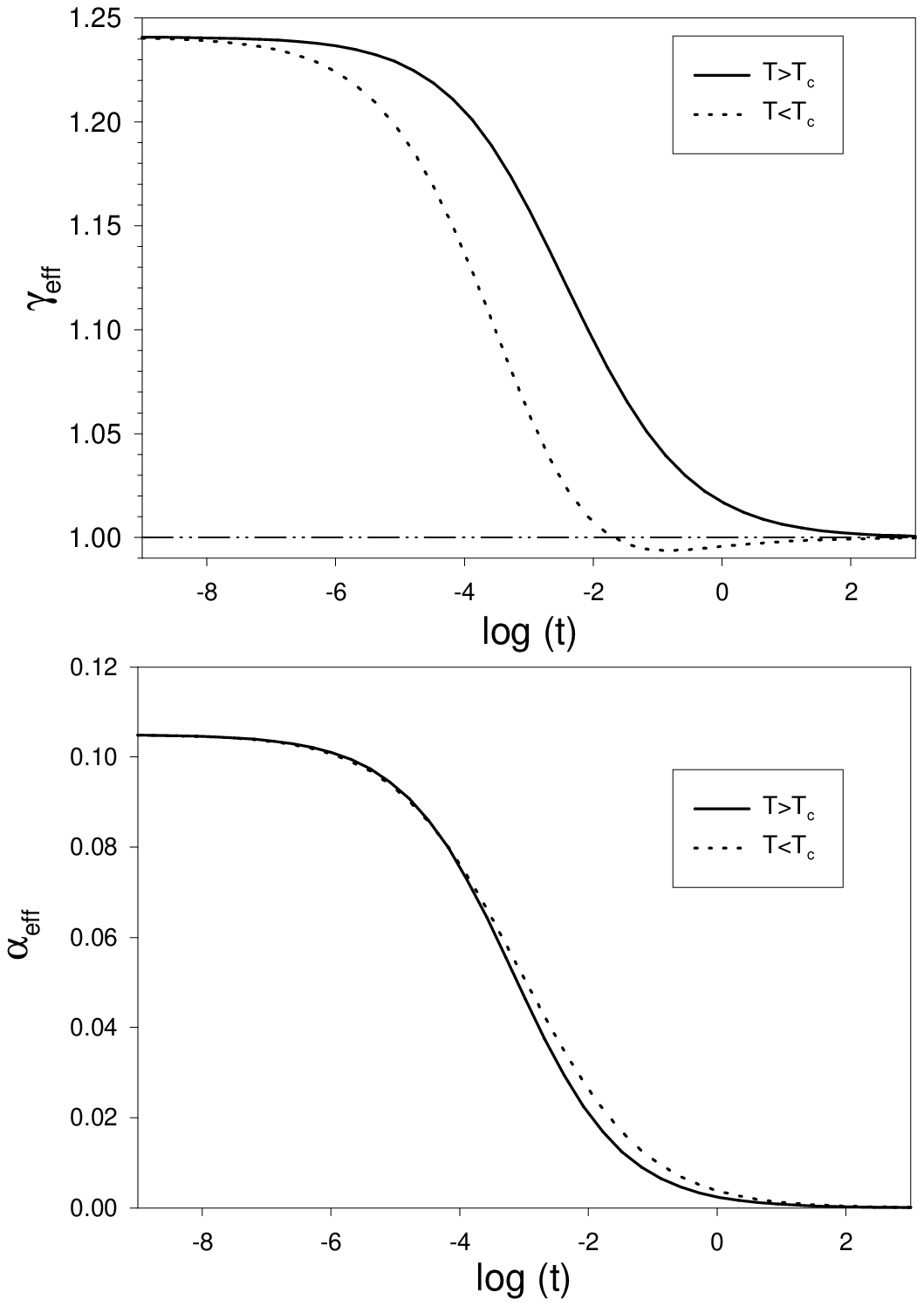} }
\caption{
Respective evolutions
(calculated from the crossover functions of tables \ref{tabhom1} and \ref
{tabinhom1}) of the effective exponents $\gamma _{eff}(t)$ and $\alpha
_{eff}(t)$ in the two phases: the homogeneous (continuous line) and the
inhomogeneous (dashed line) phases. Notice, in this latter case, the moving
down of $\gamma _{eff}(t)$ below the classical value ($=1.0$) in the regime of high
values of $t$.  This nonmonotonic feature of $\gamma _{eff}(t)$ in the inhomogeneous
phase is in agreement with Refs. \protect\cite{4073,4215}
 and has been numerically observed
in Refs.   \protect\cite{3871,4231}.%
}
\label{fig1}
\end{figure}

In the present work, the fitting procedure of step \ref{Fit} of section \ref
{remind} is performed in the entire range of values of $g\in \left]
0,g^{*}\right[ $. Consequently the entire classical-to-critical crossover
specific to the field theoretical framework is completely accounted for by
our final functions [see Eqs.\ (\ref{eq:Fdet},\ref{eq:Ddet}) and tables \ref
{tabhom1}--\ref{tabhom3}]. This is illustrated, for the Ising model ($n=1$%
), by fig. (\ref{fig1}) which displays the evolution of two effective
exponents $\gamma _{\text{eff}}\left( t\right) $ and $\alpha _{\text{eff}}\left( t\right) $, which are defined
as for example 
[see Eqs.\ (\ref{eq:notat})]: 
\begin{equation}
\gamma _{\text{eff}}\left( t\right) =-\frac{d\ln \chi \left( t\right) }{d\ln
t}  \label{eq:expeff}
\end{equation}

Fig. (\ref{fig1}) shows the effective exponents (calculated from the
crossover functions of tables \ref{tabhom1} and \ref{tabinhom1}) which
interpolate between critical and classical (mean field) values following a
form of crossover dictated by the framework of field theory. Indeed the
(massless or critical, i.e., for $t=0$) scalar field theory in three
dimensions is defined on a special trajectory of the renormalization group
(a renormalized trajectory \cite{440} [RT]) which takes its origin at
the Gaussian fixed point (characterized by classical values of the
``critical'' exponents) and joins the Wilson-Fisher fixed point (where the
critical exponents take on their critical values according to the
universality class considered). Of course, the crossover so induced is not
universal, it is specific of the framework used. In fact, strictly speaking,
only the extreme asymptotic moving away from the Wilson-Fisher fixed point
induced by small non-zero values of $t$ is universal (critical exponents and
critical amplitude combinations), even the first-correction amplitude
(defined in the close vicinity of the fixed point when $t$ is not very
small) is not universal. For example, in the present work, a specific
definite sign of the first correction amplitude is imposed due to the RT
chosen, however in actual systems that kind of correction may well be of the
opposite sign and even absent \cite{38}. Fortunately, nonuniversal does
not mean necessarily absent in actual critical behaviors. It may well occur
that actual systems (or models) display, more or less partially, the kind of crossover
calculated from field theory \cite{4627,4073,4215,4323,3869,4355}. See section \ref{USEOF} for a
practical use of the crossover functions.

Let us now give some technical informations on the fitting procedure of step 
\ref{Fit} of section \ref{remind} that must be applied two times for each
quantity considered because of the two error bounds ``$\max $'' and ``$\min $%
'' (see sections \ref{remind} and \ref{criteria} for the meaning of these
bounds).

In order to analytically reproduce the functions calculated point by point,
we use the following general form \cite{732}: 
\begin{equation}
F\left( t\right) =Z\,t^{e}\prod_{i=1}^{K}\left( 1+X_{i}t^{D(t)}\right)
^{Y_{i}}+X_{6}  \label{eq:Fdet}
\end{equation}
in which $K$ is the maximum number of factors (in a preliminary work \cite
{732} we had $K=3$, in the present work $K$ can be as large as $5$), and
with: 
\begin{equation}
D\left( t\right) =\Delta -1+\frac{S_{1}\sqrt{t}+1}{S_{2}\sqrt{t}+1}
\label{eq:Ddet}
\end{equation}
in which $\Delta $ is the correction exponent. We have adjusted each of the
parameters $Z$, $e$, $\left\{ X_{i},Y_{i}\right\} $ $(i=1,\cdots ,K)$, $%
S_{1} $, $S_{2}$, $X_{6}$ and $\Delta $ so as to fit the discretized
evolutions of the quantities considered ($\xi $, $\chi $, $C$, $M_{S}$) as
continuous functions of the temperature-like variable $t$ in the range $t\in
\left[ 10^{-17},10^{14}\right] $. Of course, there are some external
constraints on the values of these parameters which facilitate their
adjustment:

\begin{enumerate}
\item  The exponents $e$ and $\Delta $ must take on values already known
from the resummation of the corresponding elementary series.

\item  The amplitude $Z$ is easily determined with few points corresponding
to very small values of $t$.

\item  To make it easy to get a close reproduction of the crossover towards
the classical behavior when $t\rightarrow \infty $, there are constraints:

\begin{itemize}
\item on $S_{1}$, so that we have (see, for example, Eqs.\ (A9, A10) of S.
Caracciolo et al. \cite{5052}): 
\begin{equation}
D\left( t\right) \stackrel{t\rightarrow \infty }{{\rightarrow }}1/2
\label{deltacla}
\end{equation}
this leads to: 
\begin{equation}
S_{1}=S_{2}\left( \frac{3}{2}-\Delta \right)   \label{s}
\end{equation}

\item  on one of the couple $\left\{ X_{i},Y_{i}\right\} $'s by imposing
that a known classical behavior is reached in the limit $t\rightarrow \infty 
$ then it comes: 
\begin{eqnarray}
e+\frac{1}{2}\sum_{i=1}^{K}Y_{i}=e_{c}  \label{expocla} \\
Z\prod_{i=1}^{K}\left( X_{i}\right) ^{Y_{i}}=Z_{c}
\end{eqnarray}
with $e_{c}$ and $Z_{c}$ the classical values of the critical exponents and
amplitude respectively. This leads to the constraints for one of the $%
\left\{ X_{i},Y_{i}\right\} $'s: 
\begin{eqnarray}
Y_{i_{0}}=2\left( e_{c}-e\right) -\sum_{i\neq i_{0}}Y_{i}  \label{cond1surY}
\\
X_{i_{0}}=\left[ \frac{Z_{c}}{Z}\prod_{i\neq i_{0}}\left( X_{i}\right)
^{-Y_{i}}\right] ^{1/Y_{i_{0}}}
\end{eqnarray}
with the classical values $e_{c}=1$, $\frac{1}{2}$, $\frac{1}{2}$ or $0$ and 
$Z_{c}=2,$ $1,$ $\sqrt{6}$ or $B_{c}-X_{6}$ for respectively the
susceptibility, the correlation length, the coexistence curve and the
specific heat [$X_{6}$ is the additive critical part of the specific heat
and $B_{c}$ its classical value; $B_{c}=3$ in the inhomogeneous phase ($%
T<T_{c}$), while $B_{c}=0$ in the homogeneous phase ($T>T_{c}$)].
\end{itemize}
\end{enumerate}

With the above prescriptions, we have been able to reproduce the original
calculated points with a maximum (local in $t$) of relative deviation less
than $10^{-4}$ (in the worst case and for a limited number of functions
especially in the inhomogeneous phase). However, globally (mean value of the
local deviations over the entire range of $t$), the adjustment is much
better for all the quantities.

The results of those adjustments to the discrete calculated points are given
in tables \ref{tabhom1}--\ref{tabhom3}.

We emphasize that the large number of digits displayed in the tables lays no
claim to a better accuracy than in the work of GZ, it is simply required to
obtain a careful fit of the crossover functions to the discontinuous points
primarily calculated from the available perturbative series.

\subsubsection{The resummation criteria\label{criteria}}

In our preceding work \cite{733,736}, the resummation criteria of step \ref
{Leroy} of section \ref{remind} which gave the bounds ``$\max $'' and ``$%
\min $'' were not chosen so as to closely reproduce the uncertainty of the
(at that time up-to-date) estimates of universal asymptotic critical
quantities (exponents and amplitude combinations). They simply proceeded
from a primary analysis of the convergences of the elementary series [i.e., $%
\gamma \left( g\right) $, $\nu \left( g\right) $, etc$\ldots $ ] resulting
from the (unique) resummation technique considered. This makes a notable
difference because a given function brings several elements into play [see,
e.g., Eq.\ (\ref{eq:integrale})] introducing a possible frustration of the
individual resummation criteria. Moreover, when one determines the error bar
for an individual quantity, one often rounds it up because several
resummation methods may have been considered yielding answers slightly
different from each other. Since the various asymptotic critical behaviors
of the functions of interest ($\chi \left( t\right) $, etc$\ldots $) result
from the combination of a small number of elementary series
\cite{foot1}
(namely: $\gamma \left( g\right) $, $\nu \left( g\right) $, $W\left(
g\right) $ and a few amplitude functions), the individual criteria were
combined in our preceding work \cite{733,736} so as to provide an envelope
of the error accounting automatically for correlations (frustrations)
between the error bounds. This has induced some underestimation of the
errors when the universal critical exponents or amplitude combinations were
(re)-considered from the final expressions of the functions compared to
their independent estimates.

The spirit of the present work is different. We have constrained the
resummation criteria of the elementary series so as to get as closely as
possible the GZ estimates for the universal quantities despite
the possible frustrations of the error bounds mentioned above. Thus we have
taken into account the extensions up to seven loops of the series for the
critical exponents given by Murray and Nickel \cite{312}. For the reasons
indicated just above, and also because the error estimates of the amplitude
combinations of GZ are deduced from the analysis of the
parameter dependence in the equation of state \cite{3639} (they have not
been obtained from the direct analysis of specific series for the quantities
of interest), we have encountered some difficulties in fixing the
resummation criteria for some amplitude series (it is likely that GZ have
overestimated the error for some quantities). In addition, in doing so and
concerning the amplitudes we have introduced an imbalance between the error
estimates of the two phases.  Indeed our criteria are adjusted so as to get
universal {\em ratios} (or combinations) of amplitudes which, structurally
in the present work, express themselves as series strictly defined
in the inhomogeneous phase. On the contrary, the resummation criteria in the
homogeneous phase (for only one amplitude function) have been fixed without
constraint.
Consequently, the resulting error estimates of the correction amplitudes that
we presently obtain are larger than in the previous work of refs. \cite{733,736} and notably in the
inhomogeneous phase case; they are presumably overestimated (see below and part \ref{Error}).

\begin{figure}
   \centerline{ \epsfxsize=12in \epsfbox[-150 200 750 700]{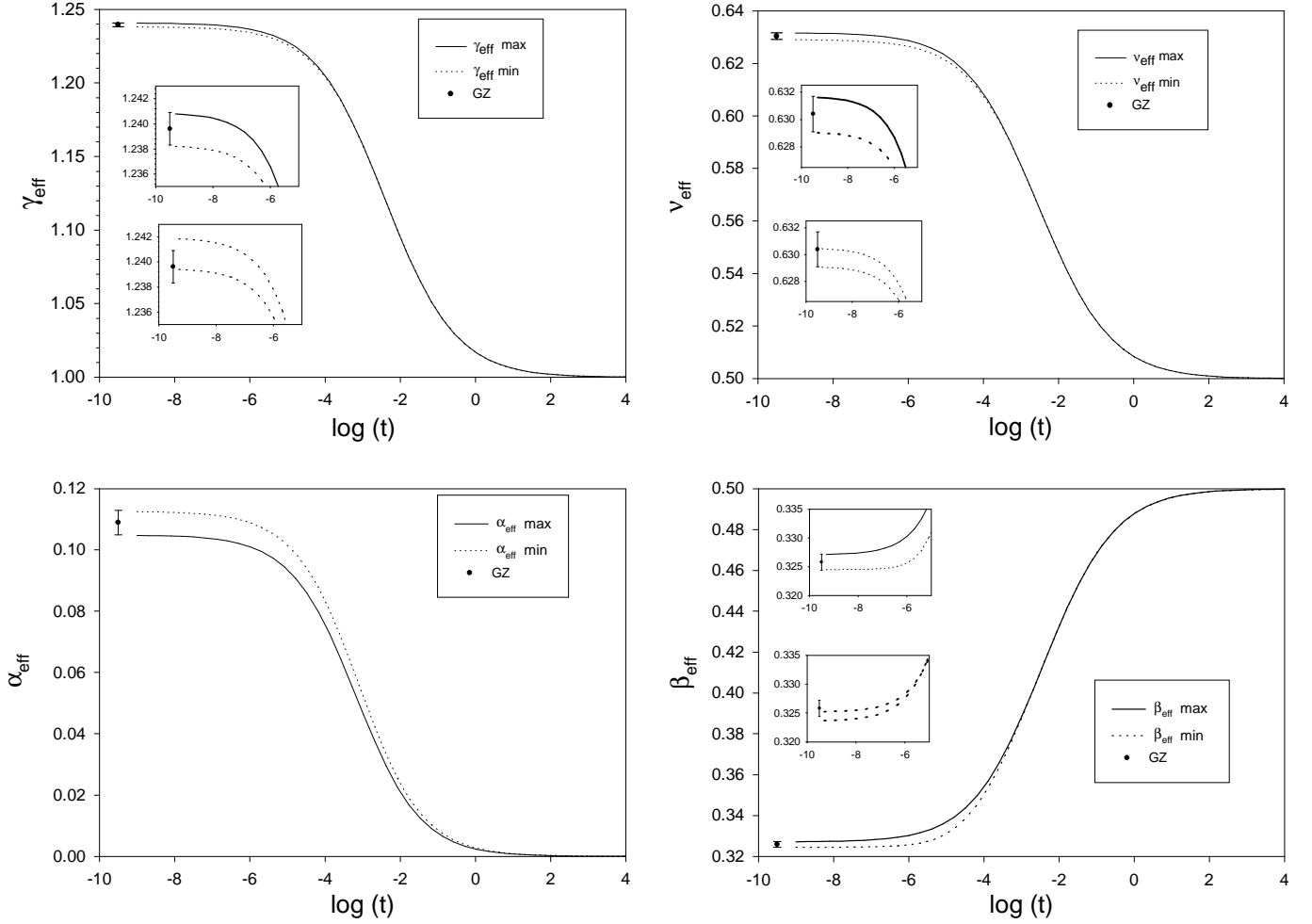} }
\caption{
The two bounds ``max''
(continuous lines) and ``min'' (short-dashed line) have been determined so
as to reproduce as closely as possible the GZ estimates%
. This is illustrated here with the effective exponents calculated from the
crossover functions for $n=1$ determined in the present work (see tables \ref
{tabhom1} and \ref{tabinhom1}). For each exponent, a partial magnification (A)
of the critical region is provided to show the agreement with the GZ estimates. A similar partial
magnification (B) is also provided to show the difference with our preceding
work of \protect\cite{733,736} (long-dashed lines). (For other values of $n$, see table \ref{tabCritExpo}.) }
\label{fig2}
\end{figure}

\begin{figure}
   \centerline{ \epsfxsize=5in \epsfbox{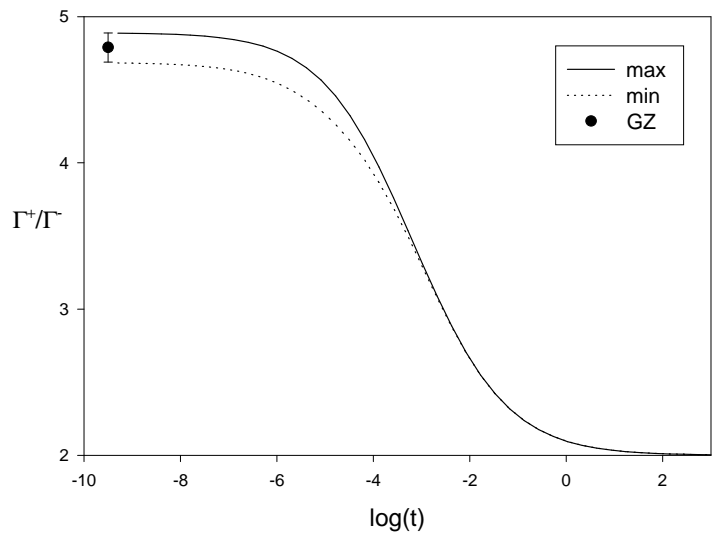} }
\caption{
The two bounds ``max''
(continuous line) and ``min'' (short-dashed line), determined so as to
reproduce as closely as possible the GZ estimates,
account also for the universal amplitude combinations. This is illustrated
here with the ratio $\Gamma ^{+}/\Gamma ^{-}$ [see Eq.\ (\ref{eq:notatb})]
calculated from the crossover functions for $n=1$ determined in the present
work (see tables \ref{tabhom1},  \ref{tabinhom1} and \ref{tabUnivAmpl}). 
The illustration could have been made  as well with the two other
universal quantities of table \ref{tabUnivAmpl}.}
\label{fig3}
\end{figure}

Table \ref{tabCritExpo} shows our estimates of the critical exponents
(resulting from our resummation criteria) compared to the GZ estimates. One
may observe some very small differences due to the fact that, in the present
work, the scaling relations are automatically satisfied for each bound ``$%
\max $'' or ``$\min $'' (see step \ref{Leroy} of section \ref{remind}) while
only the central values of the GZ exponent estimates satisfy the scaling
relations (the apparent errors for $\gamma ,\nu ,\beta $ have been
determined independently \cite{4211}). Table \ref{tabCritExpo} shows how
much the respective estimates meet the scaling relations in both cases.
Tables \ref{tabUnivAmpl} and \ref{tabUnivAmpl2} display the values of the
universal combinations of leading critical amplitudes as they are accounted
for by our crossover functions. The degree of agreement with GZ is
graphically illustrated by figs (\ref{fig2}, \ref{fig3}).

From tables \ref{tabhom1}--\ref{tabhom3} one may observe that our bounds on
the correction exponent $\Delta $ differ from GZ. This is due to the
correlation of errors mentioned above. Indeed, we have never considered $%
\Delta $ as an independent constituent of the asymptotic critical behavior.
Instead it has been (numerically) deduced from the resummation criteria
associated with the elementary series $\nu \left( g\right) $ and $W\left(
g\right) $ because of the definition $\Delta =\omega \nu $ with: 
\begin{eqnarray}
\omega &=&\left. \frac{dW\left( g\right) }{dg}\right| _{g=g^{*}} \\
\nu &=&\nu \left( g^{*}\right)
\end{eqnarray}

The resummation criteria for the elementary series $W\left( g\right) $ have
been chosen so as to yield estimates on the bounds on $g^{*}$ very close to
those of GZ (see table \ref{tabPtFix}).

Similarly, the present determinations of (and the uncertainties on) the
first correction-to-scaling terms displayed in tables \ref{tabUnivCorrAmpl1}
and \ref{tabUnivCorrAmpl2} differ from our preceding work \cite{733,736}
essentially because of our systematic account for the up-to-date estimates
of the leading amplitude combinations (see tables \ref{tabUnivAmpl} and \ref
{tabUnivAmpl2}). Notice the likely unrealistic smallness of the correction amplitude $a_M$
in the case ``min''. This confirms our probable overestimation of the error on the correction terms
(see part \ref{Error}).

It is worth indicating that the values displayed in tables 
\ref{tabUnivCorrAmpl1} and \ref{tabUnivCorrAmpl2} are not obtained from the
crossover functions of tables \ref{tabhom1}--\ref{tabhom3} by simply using
the expression [see Eq.\ (\ref{eq:Fdet})]:

\begin{equation}
a_{F}=\sum_{i=1}^{K}X_{i}Y_{i} 
\end{equation}
which would be the right expression if the correction exponent in Eq.\ (\ref
{eq:Fdet}) was the actual correction exponent $\Delta $ instead of the
effective exponent $D\left( t\right) $ of Eq.\ (\ref{eq:Ddet}). To get the
values displayed in tables \ref{tabUnivCorrAmpl1} and \ref{tabUnivCorrAmpl2}
we have made specific fits of the functions $F(t)$ of Eq.\ (\ref{eq:Fdet}) to
the theoretical points with $D\left( t\right) =\Delta $ in ranges of values
of $t<10^{-2}$ (as in the preceding work \cite{733,736}).

\section{Practical use of the crossover functions\label{USEOF}}

As already said above, the structural form of the classical-to-critical
crossover that we produce here is not universal, it is peculiar to the field
theoretical framework which corresponds to having performed a limit (the
continuum limit) in the renormalization group (RG) theory
\cite{foot2}. The approximation induces the idea that,
strictly speaking, the ``nonasymptotic'' calculations would, in fact, be
only valid in the close vicinity of $T_{c}$. Hence, {\em we do not expect
our functions to reproduce the experimental data in the entire range }$t\in
\left] 0,+\infty \right[ $. However, the width ${\cal L}$ of the domain of
agreement between experiments and field theory is not universal: it could
actually be reduced (purely and simply) to the strict asymptotic critical
region (pure scaling laws) or include exclusively the first correction to
scaling, but, fortunately it may sometimes be much larger and could even
cover the entire crossover region! It is our aim to allow experimentalists
to determine the width ${\cal L}$ of the domain of agreement. Notice that,
we do not aim at determining (or providing) all the ingredients needed to
describe the variety of classical-to-critical crossovers that may be
produced by actual systems, this would be too difficult (due to the infinite
variety of nonuniversal contributions). Simply, we think our calculation
accurate enough to allow the determination of ${\cal L}$ for any system
allowing, in some sense, the determination of subclasses of universality.

As already explained \cite{732,733,736}, the comparison of the theoretical
functions with experimental data involves a very small number of adjustable
parameters:

\begin{enumerate}
\item  Non-universal global factors,

\item  The proportionality factor $\theta $ between the temperature like
scaling field $t$ and $\tau =\frac{T-T_{c}}{T_{c}}$ (neglecting higher
analytical contributions in $\tau $ which may sometimes be non-negligible 
\cite{4748} but are out of the scope of our present aim): 
\begin{equation}
t=\theta \,\left| \tau \right|   \label{eq:theta}
\end{equation}

\item  Additive regular background terms for the specific heat,

\item  Eventually $T_{c}$.
\end{enumerate}

For example the comparison with experimental measurements of the
susceptibility $\chi $ may be made as follows: 
\begin{equation}
\chi _{0}\,\chi _{th}\left( \theta \,\left| \tau \right| \right) =\chi
_{\exp }\left( \left| \tau \right| \right)  \label{eq:chirelat}
\end{equation}
in which $\chi _{\exp }\left( \left| \tau \right| \right) $ represents the
experimental data and $\chi _{th}\left( t\right) $ our function for one
\cite{foot3} of the two
bounds ``$\max $'' and ``$\min $''. One generally expects theoretically that 
$\chi _{0}$ and $\theta $ take on the same values
\cite{foot4} for the two sets of
measurements above and below $T_{c}$ provided that the range of values of $%
\tau $ is not too large. The fact that $\chi _{0}$ must be unchanged is the
consequence of the universality of the ratio $\frac{\Gamma _{0}^{+}}{\Gamma
_{0}^{-}}$ with as $\tau \rightarrow
0^{\pm }$: 
\begin{equation}
\chi _{\exp }\left( \left| \tau \right| \right) {\simeq }\Gamma _{0}^{\pm }\left| \tau \right| ^{-\gamma }\left(
1+\Gamma _{1}^{\pm }\left| \tau \right| ^{\Delta }+\cdots \right) 
\end{equation}
in which $\Gamma _{0}^{\pm }$ and $\Gamma _{1}^{\pm }$ are related to our
previous definition [Eq.\ (\ref{eq:notatb})] as follows: 
\begin{eqnarray}
\Gamma _{0}^{\pm } &=&\chi _{0}\theta ^{-\gamma }\,\Gamma ^{\pm }
\label{eq:fac0} \\
\Gamma _{1}^{\pm } &=&\theta ^{-\Delta }a_{\chi }^{\pm }  \label{eq:fac1}
\end{eqnarray}

As for the stability of $\theta $ this is because the ratio $\frac{\Gamma
_{1}^{+}}{\Gamma _{1}^{-}}$ is universal. The values of the universal
amplitude combinations included in our calculated functions are given in
tables \ref{tabUnivAmpl}, \ref{tabUnivAmpl2}, \ref{tabUnivCorrAmpl1} and \ref
{tabUnivCorrAmpl2}.

\subsection{Redefinition of the role of $\theta $}

If one introduces $\theta $ literally as in Eq.\ (\ref{eq:theta}), then the fitted
leading critical amplitude involves two adjustable parameters. This is not
very suitable. For practical use, we propose to introduce the adjustable
parameters as follows [compare with eq (\ref{eq:Fdet})]: 
\begin{eqnarray}
\chi _{\exp }^{-1}\left( \left| \tau \right| \right) &=&\chi _{0}\,\left[
Z\,\left| \tau \right| ^{\gamma }\prod_{i=1}^{K}\left(
1+X_{i}t^{D(t)}\right) ^{Y_{i}}+X_{6}\right]  \\
t &=&\theta \,\left| \tau \right|  \label{eq:RelatGen}
\end{eqnarray}
in which $\theta $ is no longer involved in the pure scaling part of the
critical behavior ($Z\,\left| \tau \right| ^{\gamma }$). So introduced, $%
\theta $ {\em is a nonuniversal parameter which exclusively controls the
magnitude of the corrections to scaling}. Hence we can progressively adjust
the theoretical functions to the data starting from the data close to the
critical point with $\theta =0$ and then introducing more and more data with 
$\theta \neq 0$ (notice that $\theta \geq 0$) up to the point where
consistency is lost.

The domain ${\cal L}$ of $\tau $ where the experimental data and the field
theory agree may involve correction-to-scaling terms higher than the first
one and this is why it is interesting to have the theoretical expression
under the form of a complete classical-to-critical crossover
\cite{foot5}.

\paragraph*{Consistency test:} If we consider a supplementary set of measurements like the specific heat
above and below $T_{c}$, then, by virtue of universality, one must obtain
again the same value for $\theta $ with a good fit in a range of values of $%
\tau $ similar to that considered with $\chi $. For the specific heat, it
comes: 
\begin{equation}
C_{0}\,C_{th}\left( \theta \,\left| \tau \right| \right) +B_{0}(\tau
)=C_{\exp }\left( \left| \tau \right| \right)  \label{eq:Crelat}
\end{equation}
in which $B_{0}(\tau )$ is an additive non critical (i.e., regular or
analytic in $\tau $) background and $C_{0}$ a nonuniversal multiplicative
factor which must be the same in the two phases.

\begin{figure}
   \centerline{ \epsfxsize=5in \epsfbox{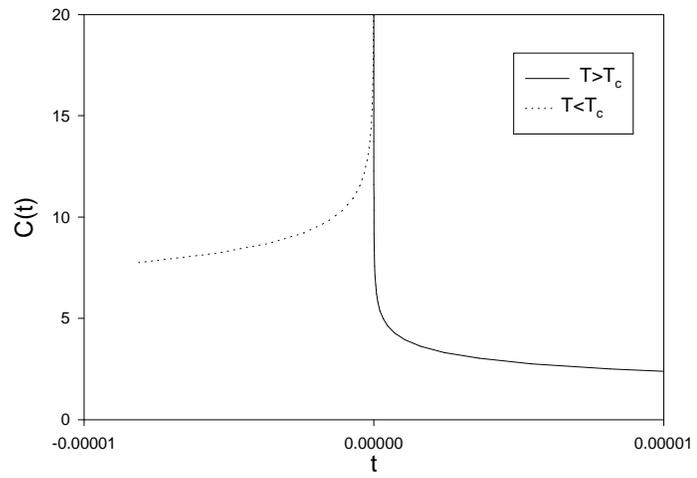} }
\caption{
The field theoretical form of the specific
heat exhibits the classical ``jump''. }
\label{fig4}
\end{figure}

Let us emphasize that the field theoretical form obtained for $C_{th}\left(
t\right) $ involves a specific critical additive background term which
reproduces the famous ``classical jump'' of the specific heat [see fig (\ref
{fig4})]. Of course, the magnitude of this jump is not universal but in the
case where an actual system would reproduce the entire classical-to-critical
crossover of field theory, then it should also exhibit this jump (up to the
global additive background $B_{0}(\tau )$ analytic in $\tau $).

If, in addition to $C$ and $\chi $, we also possess coexistence curve data,
we would have a stronger constraint since then no other adjustable parameter
would be required to fit those new data. Indeed in the relation: 
\begin{equation}
M_{0}\,M_{th}\left( \theta \,\left| \tau \right| \right) =M_{\exp }\left(
\left| \tau \right| \right)  \label{eq:Mrelat}
\end{equation}
everything is fixed since $M_{0}$ is related to $C_{0}$ and $\chi _{0}$ due
to the universal amplitude combination \cite{13} $R_{C}$ and $\theta $
must have the same value whatever the quantity considered.

If we had simultaneously also access to experimental measurements of the
correlation length $\xi $, then the constraint would be even stronger since
again the theory must agree with the data without new adjustable parameter.

In order to facilitate the use of the crossover functions displayed in table 
\ref{tabhom1}--\ref{tabhom3}, text-files of Fortran code are provided \cite
{Fortran}.

\subsection{Account for the error bounds\label{Error}}

We have accounted for the error estimates by providing two sets (``max'' and
``min'') of functions. In general the accuracy of the experimental
measurements are much smaller than in the present theoretical calculation so
that it is not very important to make a difference between the two sets of
functions. One or the other choice would provide essentially the same
quality of the adjustment in the fitting procedure.

Sometimes
accounting for the difference between the bounds ``max'' and ``min'' may
have some importance so that neither one or the other agrees with the
measurements but a mixing of the two would. In such a case we propose to
introduce the mixing via the introduction of a supplementary adjustable
parameter $E$.

Let us define a new theoretical function as follows

\begin{equation}
F_{E}\left( t\right) =\left[ F_{\max }\left( t\right) \right] ^{E}\cdot
\left[ F_{\min }\left( t\right) \right] ^{1-E} 
\end{equation}
which, regarding the definition of the effective exponents corresponds to
the linear weighting:

\begin{equation}
e_{eff}\left( t\right) =E\cdot e_{eff}^{\max }\left( t\right) +\left(
1-E\right) \cdot e_{eff}^{\min }\left( t\right) 
\end{equation}
then the introduction of the other adjustable parameters, such as $\theta $
for example, within $F_{E}\left( t\right) $ is unchanged compare to the
description given above.

As said in part \ref{criteria} it is likely that the close account of the GZ estimates
has led us to overestimate the uncertainty
on the correction terms so that it seems to us useful to provide also the reader with functions 
reproducing the complete classical-to-critical crossover according to
the resummation criteria of the previous 
(but corrected, see \cite{2376bis})
work of refs \cite{733,736} although (or rather because), this time, the error are underestimated.
This is why we provide two additional text-files of Fortran code \cite
{Fortran} corresponding to the former resummation criteria applied to the corrected series (without the
seventh
order of ref. \cite{312}). 

\paragraph*{Acknowledgements}
We thank Prof. M. Barmatz for interesting correspondence while the present
work was in progress.

\newpage

\newpage

\newpage
\widetext

\begin{table}\centering%

\caption{Numerical values of the parameters of the generic crossover function 
$F(t)$ [Eqs.\ (\ref{eq:Fdet},\ref{eq:Ddet})] corresponding to the three quantities calculated
 in the homogeneous phase ($T>T_{c}$ and for $n=1$):
the correlation length $\xi$, the susceptibility $\chi$ and
the specific heat $C$.
 For each parameter two values are provided which
correspond to the bounds ``max'' (upper line) and ``min'' (lower line) of the error treatment.
These bounds have been determined so as to reproduce as closely as possible
the error estimates of GZ on the asymptotic universal quantities (see
tables \ref{tabCritExpo} and \ref{tabUnivAmpl} and text for more details) and provide
two exclusive sets of functions $F_{\max}$ and $F_{\min}$.
 The first row of the table displays the estimate of the universal value
of the correction
exponent $\Delta$ common to all the quantities for the respective bounds ``max'' and ``min''. 
The values of the parameters have been determined
by a careful adjustment of $F(t)$ to the discrete evolution of the respective 
quantities primarily calculated by resummation of perturbation series 
using a Borel-Leroy
+ conformal mapping method.
The specific notations
of the two leading parameters $e$ (universal critical exponent) and $Z$ 
(leading critical amplitude) are recalled for each quantity [see Eq.\ (\ref{eq:notat})]. The symbol ``---'' means that the term is absent.
\label{tabhom1}}

\begin{tabular}{ccccccc}
& \multicolumn{6}{c}{$n=1$, homogeneous phase: $\Delta $ $\left\{ 
\begin{tabular}{d}
``$\max $'': 0.49862 \\ 
``$\min $'': 0.50516%
\end{tabular}
\right. $} \\ \hline
& \multicolumn{2}{c}{} & \multicolumn{2}{c}{} & \multicolumn{2}{c}{} \\ 
& \multicolumn{2}{c}{${\bf \xi }^{-1}$} & \multicolumn{2}{c}{${\bf \chi }%
^{-1}$} & \multicolumn{2}{c}{${\bf C}$} \\ 
& \multicolumn{2}{c}{} & \multicolumn{2}{c}{} & \multicolumn{2}{c}{} \\ 
\hline
\multicolumn{1}{l}{$e$} & $\nu $ & \multicolumn{1}{l}{$\left\{ 
\begin{tabular}{l}
$0.631678$ \\ 
$0.6290975$%
\end{tabular}
\right. $} & $\gamma $ & \multicolumn{1}{l}{$\left\{ 
\begin{tabular}{l}
$1.2408875$ \\ 
$1.23830$%
\end{tabular}
\right. $} & $-\alpha $ & \multicolumn{1}{l}{$\left\{ 
\begin{tabular}{l}
$-0.1049675$ \\ 
$-0.11271$%
\end{tabular}
\right. $} \\ 
\multicolumn{1}{l}{$Z$} & $\left( {\bf \xi }_{0}^{+}\right) ^{-1}$ & 
\multicolumn{1}{l}{$\left\{ 
\begin{tabular}{l}
$2.150817$ \\ 
$2.091612$%
\end{tabular}
\right. $} & $\left( {\bf \Gamma }^{+}\right) ^{-1}$ & \multicolumn{1}{l}{$%
\left\{ 
\begin{tabular}{l}
$3.75927$ \\ 
$3.660588$%
\end{tabular}
\right. $} & $\frac{A^{+}}{\alpha }$ & \multicolumn{1}{l}{$\left\{ 
\begin{tabular}{l}
$1.871810$ \\ 
$1.580112$%
\end{tabular}
\right. $} \\ 
&  &  &  &  &  &  \\ 
\multicolumn{1}{l}{$S_{1}$} &  & \multicolumn{1}{l}{$\left\{ 
\begin{tabular}{l}
$32.24878$ \\ 
$17.48596$%
\end{tabular}
\right. $} &  & \multicolumn{1}{l}{$\left\{ 
\begin{tabular}{l}
$34.05096$ \\ 
$13.38814$%
\end{tabular}
\right. $} &  & \multicolumn{1}{l}{$\left\{ 
\begin{tabular}{l}
$30.37745$ \\ 
$33.65919$%
\end{tabular}
\right. $} \\ 
\multicolumn{1}{l}{$S_{2}$} &  & \multicolumn{1}{l}{$\left\{ 
\begin{tabular}{l}
$32.20434$ \\ 
$17.57665$%
\end{tabular}
\right. $} &  & \multicolumn{1}{l}{$\left\{ 
\begin{tabular}{l}
$34.00404$ \\ 
$13.45758$%
\end{tabular}
\right. $} &  & \multicolumn{1}{l}{$\left\{ 
\begin{tabular}{l}
$30.33559$ \\ 
$33.83377$%
\end{tabular}
\right. $} \\ 
&  &  &  &  &  &  \\ 
\multicolumn{1}{l}{$X_{1}$} &  & \multicolumn{1}{l}{$\left\{ 
\begin{tabular}{l}
$11.02452$ \\ 
$10.48005$%
\end{tabular}
\right. $} &  & \multicolumn{1}{l}{$\left\{ 
\begin{tabular}{l}
$23.27915$ \\ 
$2.853295$%
\end{tabular}
\right. $} &  & \multicolumn{1}{l}{$\left\{ 
\begin{tabular}{l}
$33.31814$ \\ 
$31.94041$%
\end{tabular}
\right. $} \\ 
\multicolumn{1}{l}{$Y_{1}$} &  & \multicolumn{1}{l}{$\left\{ 
\begin{tabular}{l}
$-0.5247187$ \\ 
$-0.1283214$%
\end{tabular}
\right. $} &  & \multicolumn{1}{l}{$\left\{ 
\begin{tabular}{l}
$-0.31016527$ \\ 
$-2.547260\times 10^{-2}$%
\end{tabular}
\right. $} &  & \multicolumn{1}{l}{$\left\{ 
\begin{tabular}{l}
$3.476590$ \\ 
$0.2200185$%
\end{tabular}
\right. $} \\ 
\multicolumn{1}{l}{$X_{2}$} &  & \multicolumn{1}{l}{$\left\{ 
\begin{tabular}{l}
$10.41513$ \\ 
$28.75634$%
\end{tabular}
\right. $} &  & \multicolumn{1}{l}{$\left\{ 
\begin{tabular}{l}
$1.257832$ \\ 
$11.51061$%
\end{tabular}
\right. $} &  & \multicolumn{1}{l}{$\left\{ 
\begin{tabular}{l}
$9.400643$ \\ 
$7.017899$%
\end{tabular}
\right. $} \\ 
\multicolumn{1}{l}{$Y_{2}$} &  & \multicolumn{1}{l}{$\left\{ 
\begin{tabular}{l}
$0.3775152$ \\ 
$-9.269701\times 10^{-2}$%
\end{tabular}
\right. $} &  & \multicolumn{1}{l}{$\left\{ 
\begin{tabular}{l}
$-8.204163\times 10^{-3}$ \\ 
$-0.2766008$%
\end{tabular}
\right. $} &  & \multicolumn{1}{l}{$\left\{ 
\begin{tabular}{l}
$-8.344217\times 10^{-3}$ \\ 
$-9.616869\times 10^{-3}$%
\end{tabular}
\right. $} \\ 
\multicolumn{1}{l}{$X_{3}$} &  & \multicolumn{1}{l}{$\left\{ 
\begin{tabular}{l}
$2.315848$ \\ 
$2.014284$%
\end{tabular}
\right. $} &  & \multicolumn{1}{l}{$\left\{ 
\begin{tabular}{l}
$8.313963$ \\ 
$30.25994$%
\end{tabular}
\right. $} &  & \multicolumn{1}{l}{$\left\{ 
\begin{tabular}{l}
$33.06508$ \\ 
$0.2462918$%
\end{tabular}
\right. $} \\ 
\multicolumn{1}{l}{$Y_{3}$} &  & \multicolumn{1}{l}{$\left\{ 
\begin{tabular}{l}
$-1.307939\times 10^{-2}$ \\ 
$-6.897436\times 10^{-3}$%
\end{tabular}
\right. $} &  & \multicolumn{1}{l}{$\left\{ 
\begin{tabular}{l}
$-0.1634056$ \\ 
$-0.1745266$%
\end{tabular}
\right. $} &  & \multicolumn{1}{l}{$\left\{ 
\begin{tabular}{l}
$-3.258311$ \\ 
$-7.002609\times 10^{-5}$%
\end{tabular}
\right. $} \\ 
\multicolumn{1}{l}{$X_{4}$} &  & \multicolumn{1}{l}{$\left\{ 
\begin{tabular}{l}
$39.95028$ \\ 
$53.07716$%
\end{tabular}
\right. $} &  & $\text{---}$ &  & \multicolumn{1}{l}{$\left\{ 
\begin{tabular}{l}
$\text{---}$ \\ 
$76.39366$%
\end{tabular}
\right. $} \\ 
$Y_{4}$ &  & \multicolumn{1}{l}{$\left\{ 
\begin{tabular}{l}
$-0.1030731$ \\ 
$-3.027917\times 10^{-2}$%
\end{tabular}
\right. $} &  & $\text{---}$ &  & $\left\{ \text{%
\begin{tabular}{l}
$\text{---}$ \\ 
$1.508835\times 10^{-2}$%
\end{tabular}
}\right. $ \\ 
&  & \multicolumn{1}{l}{} &  &  &  & \multicolumn{1}{l}{} \\ 
\multicolumn{1}{l}{$X_{6}$} &  & $\text{---}$ &  & --- &  & 
\multicolumn{1}{l}{$\left\{ 
\begin{tabular}{l}
$-4.048544$ \\ 
$-3.548035$%
\end{tabular}
\right. $}
\end{tabular}

\end{table}



\begin{table}\centering

\caption{Same as table \ref{tabhom1} for the coexistence curve
 $M_S$, the susceptiblity $\chi$ and the specific heat $C$ calculated 
in the inhomogeneous phase  ($T<T_{c}$ and for $n=1$). Notice that for each bound,
the critical exponents $\gamma$, $\alpha$ and the subcritical exponent $\Delta$
take on the same values as in table  \ref{tabhom1} (as it must according to the theory). 
An indication on the accuracy of the present work is provided by 
the parameter $X_6$ (the critical background of the specific heat) which should take on
the same value as in table   \ref{tabhom1} and differs slightly for the bound ``max''.
\label{tabinhom1}}

\begin{tabular}{ccccccc}
& \multicolumn{6}{c}{$n=1$, inhomogeneous phase: $\Delta $ $\left\{ 
\begin{tabular}{l}
``$\max $'': $0.49862$ \\ 
``$\min $'': $0.50516$%
\end{tabular}
\right. $} \\ \hline
& \multicolumn{2}{c}{} & \multicolumn{2}{c}{} & \multicolumn{2}{c}{} \\ 
& \multicolumn{2}{c}{${\bf M}_{S}$} & \multicolumn{2}{c}{${\bf \chi }^{-1}$}
& \multicolumn{2}{c}{${\bf C}$} \\ 
& \multicolumn{2}{c}{} & \multicolumn{2}{c}{} & \multicolumn{2}{c}{} \\ 
\hline
\multicolumn{1}{l}{$e$} & $\beta $ & \multicolumn{1}{l}{$\left\{ 
\begin{tabular}{l}
$0.3270735$ \\ 
$0.3244954$%
\end{tabular}
\right. $} & $\gamma $ & \multicolumn{1}{l}{$\left\{ 
\begin{tabular}{l}
$1.2408875$ \\ 
$1.23830$%
\end{tabular}
\right. $} & $-\alpha $ & \multicolumn{1}{l}{$\left\{ 
\begin{tabular}{l}
$-0.1049675$ \\ 
$-0.11271$%
\end{tabular}
\right. $} \\ 
\multicolumn{1}{l}{$Z$} & $B$ & \multicolumn{1}{l}{$\left\{ 
\begin{tabular}{l}
$0.93804691$ \\ 
$0.93700952$%
\end{tabular}
\right. $} & $\left( {\bf \Gamma }^{-}\right) ^{-1}$ & \multicolumn{1}{l}{$%
\left\{ 
\begin{tabular}{l}
$18.386609$ \\ 
$17.160196$%
\end{tabular}
\right. $} & $\frac{A^{-}}{\alpha }$ & \multicolumn{1}{l}{$\left\{ 
\begin{tabular}{l}
$3.3664988$ \\ 
$3.0489086$%
\end{tabular}
\right. $} \\ 
&  &  &  &  &  &  \\ 
\multicolumn{1}{l}{$S_{1}$} &  & \multicolumn{1}{l}{$\left\{ 
\begin{tabular}{l}
$35.738988$ \\ 
$11.312578$%
\end{tabular}
\right. $} &  & \multicolumn{1}{l}{$\left\{ 
\begin{tabular}{l}
$2.3395295$ \\ 
$231.57911$%
\end{tabular}
\right. $} &  & \multicolumn{1}{l}{$\left\{ 
\begin{tabular}{l}
$1.6036462$ \\ 
$4.7885026$%
\end{tabular}
\right. $} \\ 
\multicolumn{1}{l}{$S_{2}$} &  & \multicolumn{1}{l}{$\left\{ 
\begin{tabular}{l}
$35.689736$ \\ 
$11.371253$%
\end{tabular}
\right. $} &  & \multicolumn{1}{l}{$\left\{ 
\begin{tabular}{l}
$2.3363054$ \\ 
$232.78025$%
\end{tabular}
\right. $} &  & \multicolumn{1}{l}{$\left\{ 
\begin{tabular}{l}
$1.6014362$ \\ 
$4.8133395$%
\end{tabular}
\right. $} \\ 
&  &  &  &  &  &  \\ 
\multicolumn{1}{l}{$X_{1}$} &  & \multicolumn{1}{l}{$\left\{ 
\begin{tabular}{l}
$303.21696$ \\ 
$241.51662$%
\end{tabular}
\right. $} &  & \multicolumn{1}{l}{$\left\{ 
\begin{tabular}{l}
$76.549557$ \\ 
$23.010983$%
\end{tabular}
\right. $} &  & \multicolumn{1}{l}{$\left\{ 
\begin{tabular}{l}
$79.538017$ \\ 
$123.82899$%
\end{tabular}
\right. $} \\ 
\multicolumn{1}{l}{$Y_{1}$} &  & \multicolumn{1}{l}{$\left\{ 
\begin{tabular}{l}
$-1.7687565\times 10^{-3}$ \\ 
$-5.7056933\times 10^{-2}$%
\end{tabular}
\right. $} &  & \multicolumn{1}{l}{$\left\{ 
\begin{tabular}{l}
$-3.4865628$ \\ 
$0.88915951$%
\end{tabular}
\right. $} &  & \multicolumn{1}{l}{$\left\{ 
\begin{tabular}{l}
$7.8643478\times 10^{-2}$ \\ 
$-0.26171236$%
\end{tabular}
\right. $} \\ 
\multicolumn{1}{l}{$X_{2}$} &  & \multicolumn{1}{l}{$\left\{ 
\begin{tabular}{l}
$9.3779630$ \\ 
$13.371447$%
\end{tabular}
\right. $} &  & \multicolumn{1}{l}{$\left\{ 
\begin{tabular}{l}
$59.838911$ \\ 
$61.975912$%
\end{tabular}
\right. $} &  & \multicolumn{1}{l}{$\left\{ 
\begin{tabular}{l}
$4.0631542\times 10^{-2}$ \\ 
$0.42099375$%
\end{tabular}
\right. $} \\ 
\multicolumn{1}{l}{$Y_{2}$} &  & \multicolumn{1}{l}{$\left\{ 
\begin{tabular}{l}
$0.17204103$ \\ 
$0.20926322$%
\end{tabular}
\right. $} &  & \multicolumn{1}{l}{$\left\{ 
\begin{tabular}{l}
$-16.395572$ \\ 
$4.0096724$%
\end{tabular}
\right. $} &  & \multicolumn{1}{l}{$\left\{ 
\begin{tabular}{l}
$9.1522424\times 10^{-5}$ \\ 
$-3.6312569\times 10^{-3}$%
\end{tabular}
\right. $} \\ 
\multicolumn{1}{l}{$X_{3}$} &  & \multicolumn{1}{l}{$\left\{ 
\begin{tabular}{l}
$1.3921229$ \\ 
$248.39869$%
\end{tabular}
\right. $} &  & \multicolumn{1}{l}{$\left\{ 
\begin{tabular}{l}
$3.6904512$ \\ 
$316.29079$%
\end{tabular}
\right. $} &  & \multicolumn{1}{l}{$\left\{ 
\begin{tabular}{l}
$16.574905$ \\ 
$10.791900$%
\end{tabular}
\right. $} \\ 
\multicolumn{1}{l}{$Y_{3}$} &  & \multicolumn{1}{l}{$\left\{ 
\begin{tabular}{l}
$6.0877711\times 10^{-3}$ \\ 
$-7.7226529\times 10^{-3}$%
\end{tabular}
\right. $} &  & \multicolumn{1}{l}{$\left\{ 
\begin{tabular}{l}
$4.7894058\times 10^{-2}$ \\ 
$-0.15361387$%
\end{tabular}
\right. $} &  & \multicolumn{1}{l}{$\left\{ 
\begin{tabular}{l}
$-0.28063252$ \\ 
$7.2072941\times 10^{-2}$%
\end{tabular}
\right. $} \\ 
\multicolumn{1}{l}{$X_{4}$} &  & \multicolumn{1}{l}{$\left\{ 
\begin{tabular}{l}
$30.597947$ \\ 
$82.917148$%
\end{tabular}
\right. $} &  & \multicolumn{1}{l}{$\left\{ 
\begin{tabular}{l}
$63.029796$ \\ 
$50.582996$%
\end{tabular}
\right. $} &  & \multicolumn{1}{l}{$\left\{ 
\begin{tabular}{l}
$14.361662$ \\ 
$86.921949$%
\end{tabular}
\right. $} \\ 
$Y_{4}$ &  & \multicolumn{1}{l}{$\left\{ 
\begin{tabular}{l}
$0.17144092$ \\ 
$0.15795660$%
\end{tabular}
\right. $} &  & \multicolumn{1}{l}{$\left\{ 
\begin{tabular}{l}
$19.215714$ \\ 
$-5.2595105$%
\end{tabular}
\right. $} &  & \multicolumn{1}{l}{$\left\{ \text{%
\begin{tabular}{l}
$0.13070258$ \\ 
$0.41499782$%
\end{tabular}
}\right. $} \\ 
$X_{5}$ &  & \multicolumn{1}{l}{$\left\{ 
\begin{tabular}{l}
$6.5064180$ \\ 
$4.7939978$%
\end{tabular}
\right. $} &  & \multicolumn{1}{l}{$\left\{ 
\begin{tabular}{l}
$9.3807398$ \\ 
$2.5909921$%
\end{tabular}
\right. $} &  & \multicolumn{1}{l}{$\left\{ 
\begin{tabular}{l}
$19.477188$ \\ 
$0.57982484$%
\end{tabular}
\right. $} \\ 
$Y_{5}$ &  & \multicolumn{1}{l}{$\left\{ 
\begin{tabular}{l}
$-1.9479626\times 10^{-3}$ \\ 
$4.8568967\times 10^{-2}$%
\end{tabular}
\right. $} &  & \multicolumn{1}{l}{$\left\{ 
\begin{tabular}{l}
$0.13675156$ \\ 
$3.7692433\times 10^{-2}$%
\end{tabular}
\right. $} &  & \multicolumn{1}{l}{$\left\{ 
\begin{tabular}{l}
$0.28112994$ \\ 
$3.6928566\times 10^{-3}$%
\end{tabular}
\right. $} \\ 
&  & \multicolumn{1}{l}{} &  & \multicolumn{1}{l}{} &  & \multicolumn{1}{l}{}
\\ 
\multicolumn{1}{l}{$X_{6}$} &  & \multicolumn{1}{l}{$\text{---}$} &  & 
\multicolumn{1}{l}{---} &  & \multicolumn{1}{l}{$\left\{ 
\begin{tabular}{l}
$-4.0481532$ \\ 
$-3.5480350$%
\end{tabular}
\right. $}
\end{tabular}

\end{table}



\newpage
\begin{table}\centering

\caption{Same as table \ref{tabhom1} for the correlation length $\xi$,
the susceptiblity $\chi$ and the specific heat $C$ calculated 
in the homogeneous phase  ($T>T_{c}$ and for $n=2$). Comparing with tables \ref{tabhom1} and \ref{tabhom3}
one may notice the correlation
of the value of the
 parameter $X_6$ (the critical background of the specific heat)
with the values of  leading amplitude $A^+/\alpha$ and of $\alpha$: when $\alpha$
vanishes,  $A^+$ and $\alpha X_6$  take on opposite values so as to transform the 
power law behavior $|t|^{-\alpha}$ into the logarithmic singularity $\ln |t|$.
\label{tabhom2}}

\begin{tabular}{ccccccc}
& \multicolumn{6}{c}{$n=2$, homogeneous phase: $\Delta $ $\left\{ 
\begin{tabular}{l}
``$\max $'': $0.52551$ \\ 
``$\min $'': $0.52986$%
\end{tabular}
\right. $} \\ \hline
& \multicolumn{2}{c}{} & \multicolumn{2}{c}{} & \multicolumn{2}{c}{} \\ 
& \multicolumn{2}{c}{${\bf \xi }^{-1}$} & \multicolumn{2}{c}{${\bf \chi }%
^{-1}$} & \multicolumn{2}{c}{${\bf C}$} \\ 
& \multicolumn{2}{c}{} & \multicolumn{2}{c}{} & \multicolumn{2}{c}{} \\ 
\hline
\multicolumn{1}{l}{$e$} & $\nu $ & \multicolumn{1}{l}{$\left\{ 
\begin{tabular}{l}
$0.67181082$ \\ 
$0.66878932$%
\end{tabular}
\right. $} & $\gamma $ & \multicolumn{1}{l}{$\left\{ 
\begin{tabular}{l}
$1.3188985$ \\ 
$1.3148952$%
\end{tabular}
\right. $} & $-\alpha $ & \multicolumn{1}{l}{$\left\{ 
\begin{tabular}{l}
$1.5440\times 10^{-2}$ \\ 
$6.370\times 10^{-3}$%
\end{tabular}
\right. $} \\ 
\multicolumn{1}{l}{$Z$} & $\left( {\bf \xi }_{0}^{+}\right) ^{-1}$ & 
\multicolumn{1}{l}{$\left\{ 
\begin{tabular}{l}
$2.6289918$ \\ 
$2.5496125$%
\end{tabular}
\right. $} & $\left( {\bf \Gamma }^{+}\right) ^{-1}$ & \multicolumn{1}{l}{$%
\left\{ 
\begin{tabular}{l}
$5.5612909$ \\ 
$5.3464216$%
\end{tabular}
\right. $} & $\frac{A^{+}}{\alpha }$ & \multicolumn{1}{l}{$\left\{ 
\begin{tabular}{l}
$-55.881907$ \\ 
$-121.13056$%
\end{tabular}
\right. $} \\ 
&  &  &  &  &  &  \\ 
\multicolumn{1}{l}{$S_{1}$} &  & \multicolumn{1}{l}{$\left\{ 
\begin{tabular}{l}
$15.963748$ \\ 
$33.474847$%
\end{tabular}
\right. $} &  & \multicolumn{1}{l}{$\left\{ 
\begin{tabular}{l}
$96.831346$ \\ 
$60.160224$%
\end{tabular}
\right. $} &  & \multicolumn{1}{l}{$\left\{ 
\begin{tabular}{l}
$4.0480920$ \\ 
$28.884078$%
\end{tabular}
\right. $} \\ 
\multicolumn{1}{l}{$S_{2}$} &  & \multicolumn{1}{l}{$\left\{ 
\begin{tabular}{l}
$16.381644$ \\ 
$34.505171$%
\end{tabular}
\right. $} &  & \multicolumn{1}{l}{$\left\{ 
\begin{tabular}{l}
$99.366178$ \\ 
$62.011900$%
\end{tabular}
\right. $} &  & \multicolumn{1}{l}{$\left\{ 
\begin{tabular}{l}
$4.1540622$ \\ 
$29.773102$%
\end{tabular}
\right. $} \\ 
&  &  &  &  &  &  \\ 
\multicolumn{1}{l}{$X_{1}$} &  & \multicolumn{1}{l}{$\left\{ 
\begin{tabular}{l}
$28.529734$ \\ 
$111.52736$%
\end{tabular}
\right. $} &  & \multicolumn{1}{l}{$\left\{ 
\begin{array}{c}
16.310867 \\ 
11.716728
\end{array}
\right. $} &  & \multicolumn{1}{l}{$\left\{ 
\begin{tabular}{l}
$1.1911602$ \\ 
$37.837491$%
\end{tabular}
\right. $} \\ 
\multicolumn{1}{l}{$Y_{1}$} &  & \multicolumn{1}{l}{$\left\{ 
\begin{tabular}{l}
$-9.0963764\times 10^{-2}$ \\ 
$-2.8491431\times 10^{-2}$%
\end{tabular}
\right. $} &  & \multicolumn{1}{l}{$\left\{ 
\begin{tabular}{l}
$-0.57358194$ \\ 
$-4.1213479\times 10^{-2}$%
\end{tabular}
\right. $} &  & \multicolumn{1}{l}{$\left\{ 
\begin{tabular}{l}
$5.5704053\times 10^{-2}$ \\ 
$3.5434706$%
\end{tabular}
\right. $} \\ 
\multicolumn{1}{l}{$X_{2}$} &  & \multicolumn{1}{l}{$\left\{ 
\begin{tabular}{l}
$9.1112497$ \\ 
$13.427180$%
\end{tabular}
\right. $} &  & \multicolumn{1}{l}{$\left\{ 
\begin{tabular}{l}
$3.6615694$ \\ 
$15.104245$%
\end{tabular}
\right. $} &  & \multicolumn{1}{l}{$\left\{ 
\begin{tabular}{l}
$1.2675164$ \\ 
$59.951524$%
\end{tabular}
\right. $} \\ 
\multicolumn{1}{l}{$Y_{2}$} &  & \multicolumn{1}{l}{$\left\{ 
\begin{tabular}{l}
$-0.22311836$ \\ 
$-15.783043$%
\end{tabular}
\right. $} &  & \multicolumn{1}{l}{$\left\{ 
\begin{tabular}{l}
$5.6950360\times 10^{-2}$ \\ 
$-0.53630510$%
\end{tabular}
\right. $} &  & \multicolumn{1}{l}{$\left\{ 
\begin{tabular}{l}
$-5.8499557\times 10^{-2}$ \\ 
$-1.7677012\times 10^{-2}$%
\end{tabular}
\right. $} \\ 
\multicolumn{1}{l}{$X_{3}$} &  & \multicolumn{1}{l}{$\left\{ 
\begin{tabular}{l}
$0.11326011$ \\ 
$24.100833$%
\end{tabular}
\right. $} &  & \multicolumn{1}{l}{$\left\{ 
\begin{tabular}{l}
$0.32669257$ \\ 
$3.1051883$%
\end{tabular}
\right. $} &  & \multicolumn{1}{l}{$\left\{ 
\begin{tabular}{l}
$27.562173$ \\ 
$1.3847300$%
\end{tabular}
\right. $} \\ 
\multicolumn{1}{l}{$Y_{3}$} &  & \multicolumn{1}{l}{$\left\{ 
\begin{tabular}{l}
$3.8347877\times 10^{-4}$ \\ 
$1.8612028\times 10^{-3}$%
\end{tabular}
\right. $} &  & \multicolumn{1}{l}{$\left\{ 
\begin{tabular}{l}
$-4.8535318\times 10^{-4}$ \\ 
$-3.8536260\times 10^{-2}$%
\end{tabular}
\right. $} &  & \multicolumn{1}{l}{$\left\{ 
\begin{tabular}{l}
$-7.7243929\times 10^{-3}$ \\ 
$2.2769060\times 10^{-4}$%
\end{tabular}
\right. $} \\ 
\multicolumn{1}{l}{$X_{4}$} &  & \multicolumn{1}{l}{$\left\{ 
\begin{tabular}{l}
$72.907613$ \\ 
$13.360107$%
\end{tabular}
\right. $} &  & \multicolumn{1}{l}{$\left\{ 
\begin{tabular}{l}
$430.16727$ \\ 
$239.95179$%
\end{tabular}
\right. $} &  & \multicolumn{1}{l}{$\left\{ 
\begin{tabular}{l}
$46.806723$ \\ 
$37.714984$%
\end{tabular}
\right. $} \\ 
$Y_{4}$ &  & \multicolumn{1}{l}{$\left\{ 
\begin{tabular}{l}
$-5.0166740\times 10^{-2}$ \\ 
$15.603262$%
\end{tabular}
\right. $} &  & \multicolumn{1}{l}{$\left\{ 
\begin{tabular}{l}
$-6.7793463\times 10^{-3}$ \\ 
$-1.3735564\times 10^{-2}$%
\end{tabular}
\right. $} &  & \multicolumn{1}{l}{$\left\{ \text{%
\begin{tabular}{l}
$-2.0360103\times 10^{-2}$ \\ 
$-3.5387612$%
\end{tabular}
}\right. $} \\ 
$X_{5}$ &  & \multicolumn{1}{l}{$\left\{ 
\begin{tabular}{l}
$10.299474$ \\ 
$7.3960558$%
\end{tabular}
\right. $} &  & $\text{---}$ &  & $\text{---}$ \\ 
$Y5$ &  & \multicolumn{1}{l}{$\left\{ 
\begin{tabular}{l}
$2.0243740\times 10^{-2}$ \\ 
$-0.13116817$%
\end{tabular}
\right. $} &  & $\text{---}$ &  & $\text{---}$ \\ 
&  &  &  &  &  &  \\ 
\multicolumn{1}{l}{$X_{6}$} &  & $\text{---}$ &  & --- &  & 
\multicolumn{1}{l}{$\left\{ 
\begin{tabular}{l}
$50.158572$ \\ 
$115.95104$%
\end{tabular}
\right. $}
\end{tabular}
\end{table}



\newpage
\begin{table}

\caption{Same as table \ref{tabhom1} for the correlation length $\xi$,
the susceptiblity $\chi$ and the specific heat $C$ calculated 
in the homogeneous phase  ($T>T_{c}$ and for $n=3$).
\label{tabhom3}}
\begin{tabular}{ccccccc}
& \multicolumn{6}{c}{$n=3$, homogeneous phase: $\Delta $ $\left\{ 
\begin{tabular}{l}
``$\max $'': 0.55227 \\ 
``$\min $'': 0.55702
\end{tabular}
\right. $} \\ \hline
& \multicolumn{2}{c}{} & \multicolumn{2}{c}{} & \multicolumn{2}{c}{} \\ 
& \multicolumn{2}{c}{${\bf \xi }^{-1}$} & \multicolumn{2}{c}{${\bf \chi }%
^{-1}$} & \multicolumn{2}{c}{${\bf C}$} \\ 
& \multicolumn{2}{c}{} & \multicolumn{2}{c}{} & \multicolumn{2}{c}{} \\ 
\hline
\multicolumn{1}{l}{$e$} & $\nu $ & \multicolumn{1}{l}{$\left\{ 
\begin{tabular}{l}
0.71090629 \\ 
0.70381062
\end{tabular}
\right. $} & $\gamma $ & \multicolumn{1}{l}{$\left\{ 
\begin{tabular}{l}
1.3946000 \\ 
1.3845100
\end{tabular}
\right. $} & $-\alpha $ & \multicolumn{1}{l}{$\left\{ 
\begin{tabular}{l}
0.132720 \\ 
0.11143582
\end{tabular}
\right. $} \\ 
\multicolumn{1}{l}{$Z$} & $\left( {\bf \xi }_{0}^{+}\right) ^{-1}$ & 
\multicolumn{1}{l}{$\left\{ 
\begin{tabular}{l}
3.1722403 \\ 
2.9632572
\end{tabular}
\right. $} & $\left( {\bf \Gamma }^{+}\right) ^{-1}$ & \multicolumn{1}{l}{$%
\left\{ 
\begin{tabular}{l}
7.9856105 \\ 
7.2687650
\end{tabular}
\right. $} & $\frac{A^{+}}{\alpha }$ & \multicolumn{1}{l}{$\left\{ 
\begin{tabular}{l}
$-$20.228436 \\ 
$-$18.976690
\end{tabular}
\right. $} \\ 
&  &  &  &  &  &  \\ 
\multicolumn{1}{l}{$S_{1}$} &  & \multicolumn{1}{l}{$\left\{ 
\begin{tabular}{l}
31.477107 \\ 
78.590543
\end{tabular}
\right. $} &  & \multicolumn{1}{l}{$\left\{ 
\begin{tabular}{l}
72.301387 \\ 
52.048362
\end{tabular}
\right. $} &  & \multicolumn{1}{l}{$\left\{ 
\begin{tabular}{l}
17.216858 \\ 
96640.818
\end{tabular}
\right. $} \\ 
\multicolumn{1}{l}{$S_{2}$} &  & \multicolumn{1}{l}{$\left\{ 
\begin{tabular}{l}
33.213159 \\ 
83.342746
\end{tabular}
\right. $} &  & \multicolumn{1}{l}{$\left\{ 
\begin{tabular}{l}
76.289014 \\ 
55.195616
\end{tabular}
\right. $} &  & \multicolumn{1}{l}{$\left\{ 
\begin{array}{c}
\text{18.166416} \\ 
\text{102484.48}
\end{array}
\right. $} \\ 
&  &  &  &  &  &  \\ 
\multicolumn{1}{l}{$X_{1}$} &  & \multicolumn{1}{l}{$\left\{ 
\begin{tabular}{l}
394.95293 \\ 
10.931307
\end{tabular}
\right. $} &  & \multicolumn{1}{l}{$\left\{ 
\begin{array}{c}
\text{13.735280} \\ 
\text{0.15890396}
\end{array}
\right. $} &  & \multicolumn{1}{l}{$\left\{ 
\begin{tabular}{l}
389.17897 \\ 
52.385639
\end{tabular}
\right. $} \\ 
\multicolumn{1}{l}{$Y_{1}$} &  & \multicolumn{1}{l}{$\left\{ 
\begin{tabular}{l}
$-$5.2545625$\times 10^{-3}$ \\ 
16.025379
\end{tabular}
\right. $} &  & \multicolumn{1}{l}{$\left\{ 
\begin{tabular}{l}
$-$0.75390860 \\ 
1.1525134$\times 10^{-3}$%
\end{tabular}
\right. $} &  & \multicolumn{1}{l}{$\left\{ 
\begin{tabular}{l}
$-$3.0498190$\times 10^{-3}$ \\ 
0.19945718
\end{tabular}
\right. $} \\ 
\multicolumn{1}{l}{$X_{2}$} &  & \multicolumn{1}{l}{$\left\{ 
\begin{tabular}{l}
0.15078920 \\ 
3.2123716
\end{tabular}
\right. $} &  & \multicolumn{1}{l}{$\left\{ 
\begin{tabular}{l}
1.3616777 \\ 
11.028433
\end{tabular}
\right. $} &  & \multicolumn{1}{l}{$\left\{ 
\begin{tabular}{l}
4.7607464$\times 10^{-2}$ \\ 
1179.5468
\end{tabular}
\right. $} \\ 
\multicolumn{1}{l}{$Y_{2}$} &  & \multicolumn{1}{l}{$\left\{ 
\begin{tabular}{l}
2.8641031$\times 10^{-3}$ \\ 
$-$0.11355993
\end{tabular}
\right. $} &  & \multicolumn{1}{l}{$\left\{ 
\begin{tabular}{l}
0.46960131 \\ 
$-$4.7495792$\times 10^{-2}$%
\end{tabular}
\right. $} &  & \multicolumn{1}{l}{$\left\{ 
\begin{tabular}{l}
3.2528813$\times 10^{-4}$ \\ 
1.3062104$\times 10^{-3}$%
\end{tabular}
\right. $} \\ 
\multicolumn{1}{l}{$X_{3}$} &  & \multicolumn{1}{l}{$\left\{ 
\begin{tabular}{l}
11.387266 \\ 
505.59521
\end{tabular}
\right. $} &  & \multicolumn{1}{l}{$\left\{ 
\begin{tabular}{l}
1.3862969 \\ 
12.528570
\end{tabular}
\right. $} &  & \multicolumn{1}{l}{$\left\{ 
\begin{tabular}{l}
12.991689 \\ 
22.441166
\end{tabular}
\right. $} \\ 
\multicolumn{1}{l}{$Y_{3}$} &  & \multicolumn{1}{l}{$\left\{ 
\begin{tabular}{l}
$-$0.35982658 \\ 
$-$6.9793726$\times 10^{-3}$%
\end{tabular}
\right. $} &  & \multicolumn{1}{l}{$\left\{ 
\begin{tabular}{l}
$-$0.49056204 \\ 
$-$0.69601028
\end{tabular}
\right. $} &  & \multicolumn{1}{l}{$\left\{ 
\begin{tabular}{l}
$-$0.16565010 \\ 
$-$0.13439992
\end{tabular}
\right. $} \\ 
\multicolumn{1}{l}{$X_{4}$} &  & \multicolumn{1}{l}{$\left\{ 
\begin{tabular}{l}
78.089588 \\ 
11.021201
\end{tabular}
\right. $} &  & \multicolumn{1}{l}{$\left\{ 
\begin{tabular}{l}
437.65747 \\ 
312.81912
\end{tabular}
\right. $} &  & \multicolumn{1}{l}{$\left\{ \text{%
\begin{tabular}{l}
65.185231 \\ 
36.340773
\end{tabular}
}\right. $} \\ 
$Y_{4}$ &  & \multicolumn{1}{l}{$\left\{ 
\begin{tabular}{l}
$-$5.6692560$\times 10^{-2}$ \\ 
$-$16.378555
\end{tabular}
\right. $} &  & \multicolumn{1}{l}{$\left\{ 
\begin{tabular}{l}
$-$1.4330672$\times 10^{-2}$ \\ 
$-$1.8816147$\times 10^{-2}$%
\end{tabular}
\right. $} &  & \multicolumn{1}{l}{$\left\{ 
\begin{tabular}{l}
$-$0.10948246 \\ 
$-$0.35127870
\end{tabular}
\right. $} \\ 
$X_{5}$ &  & \multicolumn{1}{l}{$\left\{ 
\begin{tabular}{l}
0.19582193 \\ 
2.8188361
\end{tabular}
\right. $} &  & \multicolumn{1}{l}{$\left\{ 
\begin{tabular}{l}
$\text{---}$ \\ 
0.97748668
\end{tabular}
\right. $} &  & \multicolumn{1}{l}{$\left\{ 
\begin{tabular}{l}
1.7080277 \\ 
11.698778
\end{tabular}
\right. $} \\ 
$Y5$ &  & \multicolumn{1}{l}{$\left\{ 
\begin{tabular}{l}
$-$2.9029786$\times 10^{-3}$ \\ 
6.6094313$\times 10^{-2}$%
\end{tabular}
\right. $} &  & \multicolumn{1}{l}{$\left\{ 
\begin{tabular}{l}
$\text{---}$ \\ 
$-$7.8502947$\times 10^{-3}$%
\end{tabular}
\right. $} &  & \multicolumn{1}{l}{$\left\{ 
\begin{tabular}{l}
1.2417093$\times 10^{-2}$ \\ 
6.2043595$\times 10^{-2}$%
\end{tabular}
\right. $} \\ 
&  &  &  &  &  &  \\ 
\multicolumn{1}{l}{$X_{6}$} &  & $\text{---}$ &  & --- &  & 
\multicolumn{1}{l}{$\left\{ 
\begin{tabular}{l}
8.2684338 \\ 
9.1558605
\end{tabular}
\right. $}
\end{tabular}

\end{table}



\newpage
\begin{table}
\caption{Above: Values of the (universal) critical exponents as they are accounted
for by the crossover functions defined in tables \ref{tabhom1}--\ref{tabhom3}.
The numbers given in parenthesis correspond to the GZ respective error-bound 
estimates.
Below: the scaling relations structurally satisfied for each bound of the crossover
functions defined in tables  \ref{tabhom1}--\ref{tabhom3}
 (the expected theoretical values are zero).
Due to the practical necessity of using a small number of
predefined criteria in the (unique) resummation method used,
the scaling relations are (automatically) better satisfied in the present work than
in the final upper and lower bounds of GZ (the numbers in parenthesis correspond
 to their bound estimates which have not particularly been
determined so as to satisfy 
the
scaling relations).
\label{tabCritExpo}}

\begin{tabular}{ccccc}
\multicolumn{5}{c}{critical exponent values} \\ 
$n$ & ${\bf \gamma }$ & ${\bf \nu }$ & $\alpha $ & $\beta $ \\ 
&  &  &  &  \\ \hline
\multicolumn{1}{l}{$1$} & \multicolumn{1}{l}{$
\begin{tabular}{l}
$1.2408875$ $(1.2409)$ \\ 
$1.23830$ $(1.2383)$%
\end{tabular}
$} & \multicolumn{1}{l}{$
\begin{tabular}{l}
$0.631678$ $(0.6317)$ \\ 
$0.6290975$ $(0.6291)$%
\end{tabular}
$} & \multicolumn{1}{l}{$
\begin{tabular}{l}
$0.1049675$ $(0.105)$ \\ 
$0.11271$ $(0.113)$%
\end{tabular}
$} & \multicolumn{1}{l}{$
\begin{tabular}{l}
$0.3270735$ $(0.3272)$ \\ 
$0.3244954$ $(0.3244)$%
\end{tabular}
$} \\ 
&  &  &  &  \\ 
$2$ & \multicolumn{1}{l}{$
\begin{tabular}{l}
$1.3188985$ $(1.3189)$ \\ 
$1.3148952$ $(1.3149)$%
\end{tabular}
$} & \multicolumn{1}{l}{$
\begin{tabular}{l}
$0.67181082$ $(0.6718)$ \\ 
$0.66878932$ $(0.6688)$%
\end{tabular}
$} & \multicolumn{1}{l}{$
\begin{tabular}{l}
$-0.01544$ ($-0.015$) \\ 
$-0.00637$ ($-0.007$)
\end{tabular}
$} & --- \\ 
&  &  &  &  \\ 
3 & $
\begin{tabular}{l}
1.39460 $(1.3945)$ \\ 
1.38451 $(1.3845)$%
\end{tabular}
$ & $
\begin{tabular}{l}
0.71090629 $(0.7108)$ \\ 
0.70381062 $(0.7038)$%
\end{tabular}
$ & $
\begin{tabular}{l}
\end{tabular}
\begin{tabular}{l}
$-$0.132720 ($-0.132$) \\ 
$-$0.11143582 ($-0.112$)
\end{tabular}
$ & --- \\ 
\multicolumn{5}{c}{} \\ \hline
\multicolumn{5}{c}{scaling relations (should be zero)} \\ 
& \multicolumn{2}{c}{$\left| \alpha -2+3\nu \right| $} & \multicolumn{2}{c}{$%
\left| \alpha +2\beta +{\bf \gamma -2}\right| $} \\ 
&  &  &  &  \\ \hline
$1$ & \multicolumn{2}{c}{$
\begin{tabular}{l}
$1.5\times 10^{-6}\;(1.\times 10^{-4})$ \\ 
$2.5\times 10^{-6}\;(3.\times 10^{-4})$%
\end{tabular}
$} & \multicolumn{2}{c}{$
\begin{tabular}{l}
$2.\times 10^{-6}\;(3.\times 10^{-4})$ \\ 
$8.\times 10^{-7}\;(1.\times 10^{-4})$%
\end{tabular}
$} \\ 
&  &  &  &  \\ 
$2$ & \multicolumn{2}{c}{$
\begin{tabular}{l}
$7.5\times 10^{-6}\;(4.\times 10^{-4})$ \\ 
$2.0\times 10^{-6}\;(6.\times 10^{-4})$%
\end{tabular}
$} & \multicolumn{2}{c}{---} \\ 
&  &  &  &  \\ 
3 & \multicolumn{2}{c}{$
\begin{tabular}{l}
$1.1\times 10^{-6}\;(4.\times 10^{-4})$ \\ 
$4.0\times 10^{-6}\;(6.\times 10^{-4})$%
\end{tabular}
$} & \multicolumn{2}{c}{} 
\end{tabular}

\end{table}



\begin{table}

\caption{Values of universal combinations of leading critical amplitudes for thermodynamic
quantities
combining calculations in the two phases, hence for $n=1$ only (from the crossover 
functions of tables \ref{tabhom1} and \ref{tabinhom1}).
The two superposed numbers correspond to the respective bounds ``max'' (upper line)
 and ``min'' (lower line).
In parenthesis are the respective bounds of the GZ estimates.\label{tabUnivAmpl}}

\begin{tabular}{ccccc}
$A^{+}/A^{-}$ &  & $\Gamma ^{+}/\Gamma ^{-}$ &  & $R_{C}^{+}=A^{+}\Gamma
^{+}/B^{2}$ \\ \hline
\multicolumn{1}{l}{
\begin{tabular}{l}
$0.55601$\ $(0.556)$ \\ 
$0.51826\;(0.518)$%
\end{tabular}
} &  & \multicolumn{1}{l}{
\begin{tabular}{l}
$4.89100\;(4.89)$ \\ 
$4.68783\;(4.69)$%
\end{tabular}
} &  & \multicolumn{1}{l}{
\begin{tabular}{l}
$0.05940\;(0.0594)$ \\ 
$0.05541\;(0.0554)$%
\end{tabular}
}
\end{tabular}

\end{table}



\begin{table}

\caption{ Values of the universal quantity $R_{\xi }^{+}=\xi _{0}^{+}\left( A^{+}\right) ^{1/3}$
combining calculations in the homogeneous phase only, hence for three values of $n$
(from the crossover 
functions of tables \ref{tabhom1}, \ref{tabhom2} and \ref{tabhom3}). 
The two superposed numbers correspond to the respective bounds ``max'' (upper line)
 and ``min'' (lower line).
In parenthesis are rounded forms of the same estimates.
 \label{tabUnivAmpl2}}

\begin{tabular}{cccc}
$n$ & $1$ & 2 & 3 \\ \hline
\multicolumn{1}{l}{$R_{\xi }^{+}$} & \multicolumn{1}{l}{
\begin{tabular}{l}
$0.27029$ \\ 
$0.26899$%
\end{tabular}
(0.2696$\pm 0.0007$)} & 
\begin{tabular}{l}
0.36212 \\ 
0.35974
\end{tabular}
(0.3609$\pm 0.0012$) & 
\begin{tabular}{l}
0.43813 \\ 
0.43316
\end{tabular}
(0.4357$\pm 0.0025$)
\end{tabular}

\end{table}



\begin{table}

\caption{Bounds on the fixed point values of the renormalized coupling g
defined as the zero of the Wilson function $W(g)$. The resummation criteria for $W(g)$ 
have been chosen so as to yield values of $g^*$ as close as possible to the GZ estimates given in parenthesis.
The two superposed numbers correspond to the respective bounds ``max'' (upper line)
 and ``min'' (lower line).
 \label{tabPtFix}}

\begin{tabular}{cccc}

$n$ & $1$ & 2 & 3 \\ \hline
\multicolumn{1}{l}{$g^{*}$} & \multicolumn{1}{l}{
\begin{tabular}{l}
$1.41512$\ $(1.415)$ \\ 
$1.40687$\ $(1.407)$%
\end{tabular}
} & 
\begin{tabular}{l}
$1.40602$\ $(1.406)$ \\ 
$1.40004$\ $(1.400)$%
\end{tabular}
& 
\begin{tabular}{l}
$1.39401$\ $(1.394)$ \\ 
$1.38605$\ $(1.386)$%
\end{tabular}

\end{tabular}

\end{table}



\begin{table}

\caption{Values of the universal ratios of the amplitudes of the first correction
to scaling for thermodynamic quantities combining calculations in the two phases, 
hence for $n=1$ only (not obtained from the crossover functions, see text). 
Same presentation as in table  \ref{tabUnivAmpl2}.
\label{tabUnivCorrAmpl1}}

\begin{tabular}{ccccc}
$a_{\chi }^{+}/a_{\chi }^{-}$ &  & $a_{C}^{+}/a_{C}^{-}$ &  & $a_{M}/a_{\chi
}^{+}$ \\ \hline
\multicolumn{1}{l}{
\begin{tabular}{l}
$0.243$ \\ 
$0.186$%
\end{tabular}
 (0.215$\pm 0.029$)} &  & \multicolumn{1}{l}{
\begin{tabular}{l}
$0.893$ \\ 
$1.823$%
\end{tabular}
(1.36$\pm 0.47$)} &  & \multicolumn{1}{l}{
\begin{tabular}{l}
$0.743$ \\ 
$0.048$%
\end{tabular}
(0.40$\pm 0.35$)} 
\end{tabular}

\end{table}



\begin{table}

\caption{Values of the universal ratios of the amplitudes of the first correction
to scaling for the quantities calculated in the homogeneous phase 
 (not obtained from the crossover functions, see text).
Same presentation as in table \ref{tabUnivAmpl2}.
\label{tabUnivCorrAmpl2}}

\begin{tabular}{lccc}

$n$ & 1 & 2 & 3 \\ \hline
\multicolumn{1}{c}{$a_{\xi }^{+}/a_{\chi }^{+}$} & \multicolumn{1}{l}{$%
\left\{ 
\begin{tabular}{l}
$0.699$ \\ 
$0.659$%
\end{tabular}
\right. $(0.68$\pm 0.02$)} & \multicolumn{1}{l}{
\begin{tabular}{l}
$0.642$ \\ 
$0.630$%
\end{tabular}
(0.636$\pm 0.006$)} & \multicolumn{1}{l}{
\begin{tabular}{l}
$0.625$ \\ 
$0.598$%
\end{tabular}
(0.612$\pm 0.014$)} \\ 
\multicolumn{1}{c}{$a_{C}^{+}/a_{\chi }^{+}$} & \multicolumn{1}{l}{$\left\{ 
\begin{tabular}{l}
$8.89$ \\ 
$8.43$%
\end{tabular}
\right. $(8.68$\pm 0.23$)} & \multicolumn{1}{l}{
\begin{tabular}{l}
$6.09$ \\ 
5.$84$%
\end{tabular}
(5.97$\pm 0.13$)} & \multicolumn{1}{l}{
\begin{tabular}{l}
$4.63$ \\ 
$4.52$%
\end{tabular}
(4.58$\pm 0.06$)}
\end{tabular}

\end{table}
\newpage
\appendix
\section{Computer Program in Fortran \\ for the Ising model in the two phases
\\ (up-to-date criteria)}
\begin{verbatim}
       program utdFTcross1
c************************************************
c                   up-to-date                  *
c Crossover functions up to seven loop orders   *
c from Murray-Nickel (1991) and                 *
c accounting for the recent analysis by Guida   *
c and Zinn-Justin [J Phys A31, 8103 (1998)]     *
c************************************************
c For Ising-like systems (n=1) above & below Tc *
c************************************************
c see the companion paper:                      *
C "Classical-to-critical crossovers from field  *
c theory" by C. Bagnuls and C. Bervillier       *
c************************************************
c The important subroutine is 
c fns(res,iphase,ibound,ifn,tau,theta,ikont)
c res is the return value of the subroutine
c iphase controls the phase: 
c       iphase=1 corresponds to T>Tc
c       iphase=2 corresponds to T<Tc
c ifn controls the type of function chosen:
c       ifn=1 gives the inverse correlation length for T>Tc if iphase=1
c otherwise (iphase=2) it gives the coexistence curve
c       ifn=2 gives the inverse susceptibility
c       ifn=3 gives the specific heat
c ibound controls the accuracy of the theoretical calculation:
c       ibound=1 corresponds to "max"
c       ibound=2 corresponds to "min"
c tau is |T-Tc|/Tc
c theta is the adjustable parameter which relates the scaling field t to tau:
c     (in principle t=theta*tau but for practical use we have separated the asymptotic
c      pure scaling from the correction-to-scaling contributions, see the companion paper,
c      for theta=1 one recovers the pure theoretical
c      functions of the scaling field t)
c ikont is an integer which allows to choose the nature of res
c      if ikont=0 then res gives the functions as indicated above
c      if ikont=1 then res corresponds to the effective exponent associated to the function
c          chosen according to the criteria described above. 
c All the functions are for Ising-like model (n=1)
C********************************
      implicit double precision (a-h,o-z)       
C*************************************************************
C Evolution of the functions and of the effective exponents  *
C  in terms of the pure scaling field t                      *
C  (i.e., theta=1)                                           *
C*************************************************************
C
C***
C Selection of the function
C***
      iphase=2    ! controls the phase: 1=T>Tc, 2=T<Tc
      ibound=1    ! controls the accuracy bound : 1=max, 2=min
      ifn=3       ! controls the fn: 1=xi, 2=chi, 3=c; for T<Tc, 1=coex curve
C***
C End of selection
C***
C***
C Files for saving the results
C***
      open(29,FILE='expeff.dat')
      open(30,FILE='Funct.dat')
      theta=1                    ! gives the theoretical evolution in terms of the scaling field t
      t=1.d-9
      do 30 i=1,45
      call fns(f3,iphase,ibound,ifn,t,theta,0)   !Call of the function with ikont=0
      write(30,666) t,f3
      call fns(exp2,iphase,ibound,ifn,t,theta,1)   ! Call of the effective exponent with ikont=1
      print *, t,f3,exp2
      alogt=dlog10(t)
      write(29,666) alogt,exp2     ! Ready for a Log-Lin plot
      t=t*2
30    continue
      close(29)
      close(30)
      stop   
666   format(2(G13.6,1x))
      end
C***
C Theoretical functions for n=1
C***
      SUBROUTINE fns(res,i,j,k,t,theta,ikont)
c*************************************************
c if ikont=0: res gives the function
c if ikont=1: res gives the effective exponent
c Table f contains the values of the functions' parameters.
c At the end of each set a comment allows to distinguish the
c kind of function concerned.
c For each set, the first value is the critical exponent value
c  (or minus this value for the specific heat), the second value
c  is the inverse of the critical amplitude for xi and chi, the
c  critical amplitude for aim and c, the third value is the critical
c  background which is non zero only for c.
c The two values of the subcritical exponent delta corresponding to the two 
c bounds "max" and "min" are contained in the table del
c*************************************************
      implicit double precision (a-h,o-z)
      dimension del(2),x(5),y(5)
      dimension f(15,3,2,2)
      data del/0.49862,0.50516/
      data F/0.631678,2.150817,0.,32.24878,32.20434,11.02452,
     #-0.5247187,10.41513,0.3775152,2.315848,-1.307939D-02,
     #39.95028,-1.030731D-01,0.,0.,                                       !ximax  T>Tc (corr length)
     #1.2408875,3.75927,0.,34.05096,34.00404,23.27915,-0.31016527,
     #1.257832,-8.204163D-03,8.313963,-0.1634056,0.,0.,0.,0.,             !chimax  T>Tc (suscept)
     #-.1049675,1.871810,-4.048544,30.37745,30.33559,33.31814,3.476590,
     #9.400643,-8.344217D-03,33.06508,-3.258311,0.,0.,0.,0.,              !cmax  T>Tc (specif heat)
     #0.6290975,2.091612,0.,17.48596,17.57665,10.48005,-0.1283214,
     #28.75634,-9.269701D-02,2.014284,-6.897436D-03,53.07716,
     #-3.027917D-02,0.,0.,                                                !ximin  T>Tc (corr length)
     #1.23830,3.660588,0.,13.38814,13.45758,2.853295,-2.547260D-02,
     #11.51061,-0.2766008,30.25994,-0.1745266,0.,0.,0.,0.,                !chimin  T>Tc (suscept)
     #-.11271,1.580112,-3.548035,33.65919,33.83377,31.94041,
     #0.2200185,7.017899,-9.616869D-03,0.2462918,-7.002609D-05,76.39366,
     #1.508835D-02,0.,0.,                                                 !cmin  T>Tc (specif heat)
     #0.32707350,0.93804691,0.,35.738988,35.689736,
     #303.21696,-0.17687565D-02,9.3779630,0.17204103,1.3921229,
     #0.60877711D-02,30.597947,0.17144092,6.5064180,-0.19479626D-02,      !aimax  T<Tc (coex curve)
     #1.2408875,18.386609,0.,2.3395295,2.3363054,76.549557,-3.4865628,
     #59.838911,-16.395572,3.6904512,0.47894058D-01,63.029796,
     #19.215714,9.3807398,0.13675156,                                     !chimax  T<Tc (suscept)
     #-0.10496750,3.3664988,-4.0481532,1.6036462,1.6014362,
     #79.538017,0.78643478D-01,0.40631542D-01,0.91522424D-04,16.574905,
     #-0.28063252,14.361662,0.13070258,19.477188,0.28112994,              !cmax  T<Tc (specif heat)
     #0.32449540,0.93700952,0.,11.312578,11.371253,
     #241.51662,-0.57056933D-01,13.371447,0.20926322,248.39869,
     #-0.77226529D-02,82.917148,0.15795660,4.7939978,0.48568967D-01,      !aimin  T<Tc (coex curve)
     #1.2383000,17.160196,0.,231.57911,232.78025,23.010983,
     #0.88915951,61.975912,4.0096724,316.29079,-0.15361387,
     #50.582996,-5.2595105,2.5909921,0.37692433D-01,                      !chimin  T<Tc (suscept)
     #-0.11271000,3.0489086,-3.5480350,4.7885026,4.8133395,
     #123.82899,-0.26171236,0.42099375,-0.36312569D-02,10.791900,
     #0.72072941D-01,86.921949,0.41499782,0.57982484,0.36928566D-02/      !cmin  T<Tc (specif heat)
C*********
C  definitions of constants
C*********
      d=del(j)    ! correction exponent
      e=F(1,k,j,i) ! asymptotic exponent
      z=f(2,k,j,i)  ! asymptotic amplitude 
      x6=f(3,k,j,i) ! additive critical background
      s1=f(4,k,j,i)
      s2=f(5,k,j,i)
      kmax=0
      do 1 ii=1,5
      x(ii)=f(2*ii+4,k,j,i)    ! definitions of the Xi's and Yi's
      if(x(ii).eq.0.) go to 2
      y(ii)=f(2*ii+5,k,j,i)
1     kmax=kmax+1
2     continue
      if(ikont.eq.0) then    ! calculation of the function
      res=Z*t**e
      if(theta.eq.0.) then   ! If theta=0
      res=res+x6             ! there is no correction-to-scaling
      return                 ! only the pure scaling form
      end if                 ! survives
      tt=theta*t
      D=D-1+(S1*dsqrt(tt)+1)/(S2*dsqrt(tt)+1)
      trr=(tt)**D
      do 3 kk=1,kmax
3     res=res*(1+X(kk)*trr)**Y(kk)
      res=res+X6
      return
      end if
      if(ikont.eq.1) then    ! calculation of the effective exponent
      if(theta.eq.0) then   ! If theta=0
      res=e                 ! then
      if(k.eq.3) res=-res   ! the effective exponent
      return                ! reduces to the constant 
      end if                ! critical exponent
      tt=theta*t
      D=D-1+(S1*dsqrt(tt)+1)/(S2*dsqrt(tt)+1)
      trr=(tt)**D
      Dp=(S1-S2)/(2*dsqrt(tt)*(S2*dsqrt(tt)+1)**2)
      res=0.
      do 4 kk=1,kmax
4     res=res+X(kk)*Y(kk)/(1+X(kk)*trr)
      res=res*(Dp*tt*DLOG(tt)+D)*trr
      res=res+e
      if(k.eq.3) res=-res     ! the exponent changes sign for c
      return
      end if
      print *, "Error, ikont is ",ikont, ", but must be 0 or 1."
      stop
      end
\end{verbatim}

\newpage
\section{Computer Program in Fortran \\ for the $n$-vector model ($n=1, 2, 3$) in the homogeneous phase
\\ (up-to-date criteria)}
\begin{verbatim}
      program utdFTcrossN
c************************************************
c                   up-to-date                  *
c Crossover functions up to seven loop orders   *
c from Murray-Nickel (1991) and                 *
c accounting for the recent analysis by Guida   *
c and Zinn-Justin [J Phys A31, 8103 (1998)]     *
c************************************************
c For N-vector-like systems (n=1, 2, 3)         *
c in the homogeneous phase only                 *
c************************************************
c see the companion paper:                      *
C "Classical-to-critical crossovers from field  *
c theory" by C. Bagnuls and C. Bervillier       *
c************************************************
c The important subroutine is 
c fns(res,iN,ibound,ifn,tau,theta,ikont)
c res is the return value of the subroutine
c iN controls the value of n: 
c       iN=1 corresponds to n=1
c       iN=2 corresponds to n=2
c       iN=3 corresponds to n=3
c ifn controls the type of function chosen:
c       ifn=1 gives the inverse correlation length
c       ifn=2 gives the inverse susceptibility
c       ifn=3 gives the specific heat
c ibound controls the accuracy of the theoretical calculation:
c       ibound=1 corresponds to "max"
c       ibound=2 corresponds to "min"
c tau is |T-Tc|/Tc
c theta is the adjustable parameter which relates the scaling field t to tau:
c     (in principle t=theta*tau but for practical use we have separated the asymptotic
c      pure scaling from the correction-to-scaling contributions, see the companion paper,
c      for theta=1 one recovers the pure theoretical
c      functions of the scaling field t)
c ikont is an integer which allows to choose the nature of res
c      if ikont=0 then res gives the functions as indicated above
c      if ikont=1 then res corresponds to the effective exponent associated to the function
c          chosen according to the criteria described above. 
C********************************
      implicit double precision (a-h,o-z)       
C*************************************************************
C Evolution of the functions and of the effective exponents  *
C  in terms of the pure scaling field t                      *
C  (i.e., theta=1)                                           *
C*************************************************************
C
C***
C Selection of the function
C***
      iN=2        ! controls the value of n: 1, 2, or 3
      ibound=1    ! controls the accuracy bound : 1=max, 2=min
      ifn=3       ! controls the fn: 1=xi, 2=chi, 3=c
C***
C End of selection
C***
C***
C Files for saving the results
C***
      open(29,FILE='expeff.dat')
      open(30,FILE='Funct.dat')
      theta=1                    ! gives the theoretical evolution in terms of the scaling field t
      t=1.d-9
      do 30 i=1,45
      call fns(f3,iN,ibound,ifn,t,theta,0)   !Call of the function with ikont=0
      write(30,666) t,f3
      call fns(exp2,iN,ibound,ifn,t,theta,1)   ! Call of the effective exponent with ikont=1
      print *, t,f3,exp2
      alogt=dlog10(t)
      write(29,666) alogt,exp2     ! Ready for a Log-Lin plot
      t=t*2
30    continue
      close(29)
      close(30)
      stop   
666   format(2(G13.6,1x))
      end
C***
C Theoretical functions for n-vector-like systems
C***
      SUBROUTINE fns(res,i,j,k,t,theta,ikont)
c*************************************************
c if ikont=0: res gives the function
c if ikont=1: res gives the effective exponent
c Table f contains the values of the functions' parameters.
c At the end of each set a comment allows to distinguish the
c kind of function concerned.
c For each set, the first value is the critical exponent value
c  (or minus this value for the specific heat), the second value
c  is the inverse of the critical amplitude for xi and chi, the
c  critical amplitude for c, the third value is the critical
c  background which is non zero only for c.
c The six values of the subcritical exponent delta corresponding to the two 
c bounds "max" and "min" for each value of n are contained in the table del
c*************************************************
      implicit double precision (a-h,o-z)
      dimension del(2,3),x(5),y(5)
      dimension f(15,3,2,3)
      data del/0.49862,0.50516,0.52551,0.52986,0.55227,0.55702/
      data F/0.631678,2.150817,0.,32.24878,32.20434,11.02452,
     #-0.5247187,10.41513,0.3775152,2.315848,-1.307939D-02,
     #39.95028,-1.030731D-01,0.,0.,                                       !ximax n=1 (corr length)
     #1.2408875,3.75927,0.,34.05096,34.00404,23.27915,-0.31016527,
     #1.257832,-8.204163D-03,8.313963,-0.1634056,0.,0.,0.,0.,             !chimax n=1 (suscept)
     #-.1049675,1.871810,-4.048544,30.37745,30.33559,33.31814,3.476590,
     #9.400643,-8.344217D-03,33.06508,-3.258311,0.,0.,0.,0.,              !cmax n=1 (specif heat)
     #0.6290975,2.091612,0.,17.48596,17.57665,10.48005,-0.1283214,
     #28.75634,-9.269701D-02,2.014284,-6.897436D-03,53.07716,
     #-3.027917D-02,0.,0.,                                                !ximin  n=1 (corr length)
     #1.23830,3.660588,0.,13.38814,13.45758,2.853295,-2.547260D-02,
     #11.51061,-0.2766008,30.25994,-0.1745266,0.,0.,0.,0.,                !chimin  n=1 (suscept)
     #-.11271,1.580112,-3.548035,33.65919,33.83377,31.94041,
     #0.2200185,7.017899,-9.616869D-03,0.2462918,-7.002609D-05,76.39366,
     #1.508835D-02,0.,0.,                                                 !cmin  n=1 (specif heat)
     #0.67181082,2.6289918,0.,15.963748,16.381644,
     #28.529734,-0.90963764D-01,9.1112497,-0.22311836,0.11326011,
     #0.38347877D-03,72.907613,-0.50166740D-01,10.299474,0.20243740D-01,  !ximax n=2 (corr length)
     #1.3188985,5.5612909,0.,96.831346,99.366178,
     #16.310867,-0.57358194,3.6615694,-0.56950360D-01,0.32669257,
     #-0.48535318D-03,430.16727,-0.67793463D-02,0.,0.,                    !chimax n=2 (suscept)
     #0.15440000D-01,-55.881907,50.158572,4.0480920,4.1540622,
     #1.1911602,0.55704053D-01,1.2675164,-0.58499557D-01,27.562173,
     #-0.77243929D-02,46.806723,-0.20360103D-01,0.,0.,                    !cmax n=2 (specif heat)
     #0.66878932,2.5496125,0.,33.474847,34.505171,
     #111.52736,-0.28491431D-01,13.427180,-15.783043,24.100833,
     #0.18612028D-02,13.360107,15.603262,7.3960558,-0.13116817,           !ximin  n=2 (corr length)
     #1.3148952,5.3464216,0.,60.160224,62.011900,
     #11.716728,-0.41213479D-01,15.104245,-0.53630510,3.1051883,
     #-0.38536260D-01,239.95179,-0.13735564D-01,0.,0.,                    !chimin  n=2 (suscept)
     #0.63700000D-02,-121.13056,115.95104,28.884078,29.773102,
     #37.837491,3.5434706,59.951524,-0.17677012D-01,1.3847300,
     #0.22769060D-03,37.714984,-3.5387612,0.,0.,                          !cmin  n=2 (specif heat)
     #0.71090629,3.1722403,0.,31.477107,33.213159,
     #394.95293,-0.52545625D-02,0.15078920,0.28641031D-02,11.387266,
     #-0.35982658,78.089588,-0.56692560D-01,0.19582193,-0.29029786D-02,   !ximax n=3 (corr length)
     #1.3946000,7.9856105,0.,72.301387,76.289014,
     #13.735280,-0.75390860,1.3616777,0.46960131,1.3862969,
     #-0.49056204,437.65747,-0.14330672D-01,0.,0.,                        !chimax n=3 (suscept)
     #0.13272000,-20.228436,8.2684338,17.216858,18.166416,
     #389.17897,-0.30498190D-02,0.47607464D-01,0.32528813D-03,12.991689,
     #-0.16565010,65.185231,-0.10948246,1.7080277,0.12417093D-01,         !cmax n=3 (specif heat)
     #0.70381062,2.9632572,0.,78.590543,83.342746,
     #10.931307,16.025379,3.2123716,-0.11355993,505.59521,
     #-0.69793726D-02,11.021201,-16.378555,2.8188361,0.66094313D-01,      !ximin  n=3 (corr length)
     #1.3845100,7.2687650,0.,52.048362,55.195616,
     #0.15890396,0.11525134D-02,11.028433,-0.47495792D-01,12.528570,
     #-0.69601028,312.81912,-0.18816147D-01,0.97748668,-0.78502947D-02,   !chimin  n=3 (suscept)
     #0.11143582,-18.976690,9.1558605,96640.818,102484.48,
     #52.385639,0.19945718,1179.5468,0.13062104D-02,22.441166,
     #-0.13439992,36.340773,-0.35127870,11.698778,0.62043595D-01/         !cmin  n=3 (suscept)
C*********
C  definitions of constants
C*********
      if(i.gt.3.or.i.lt.1) then
      print *, "Error, n= ",i, "is not available. Sorry. (0<n<4)"
      stop
      end if
      d=del(j,i)    ! correction exponent
      e=F(1,k,j,i) ! asymptotic exponent
      z=f(2,k,j,i)  ! asymptotic amplitude 
      x6=f(3,k,j,i) ! additive critical background
      s1=f(4,k,j,i)
      s2=f(5,k,j,i)
      kmax=0
      do 1 ii=1,5
      x(ii)=f(2*ii+4,k,j,i)    ! definitions of the Xi's and Yi's
      if(x(ii).eq.0.) go to 2
      y(ii)=f(2*ii+5,k,j,i)
1     kmax=kmax+1
2     continue
      if(ikont.eq.0) then    ! calculation of the function
      res=Z*t**e
      if(theta.eq.0.) then   ! If theta=0
      res=res+x6             ! there is no correction-to-scaling
      return                 ! only the pure scaling form
      end if                 ! survives
      tt=theta*t
      D=D-1+(S1*dsqrt(tt)+1)/(S2*dsqrt(tt)+1)
      trr=(tt)**D
      do 3 kk=1,kmax
3     res=res*(1+X(kk)*trr)**Y(kk)
      res=res+X6
      return
      end if
      if(ikont.eq.1) then    ! calculation of the effective exponent
      if(theta.eq.0) then   ! If theta=0
      res=e                 ! then
      if(k.eq.3) res=-res  ! the effective exponent
      return                ! reduces to the constant 
      end if                ! critical exponent
      tt=theta*t
      D=D-1+(S1*dsqrt(tt)+1)/(S2*dsqrt(tt)+1)
      trr=(tt)**D
      Dp=(S1-S2)/(2*dsqrt(tt)*(S2*dsqrt(tt)+1)**2)
      res=0.
      do 4 kk=1,kmax
4     res=res+X(kk)*Y(kk)/(1+X(kk)*trr)
      res=res*(Dp*tt*DLOG(tt)+D)*trr
      res=res+e
      if(k.eq.3) res=-res  ! the exponent changes sign for c
      return
      end if
      print *, "Error, ikont is ",ikont, ", but must be 0 or 1."
      stop
      end
\end{verbatim}

\newpage
\section{Computer Program in Fortran \\ for the Ising model in the two phases
\\ (former criteria, corrected series)}
\begin{verbatim}
      program FTcross1
c************************************************
c            former corrected                   *
c Crossover functions up to six loop orders     *
c after correction of the errors for the        *
c inhomogeneous phase                           *
c************************************************
c For Ising-like systems (n=1) above & below Tc *
c************************************************
c see the companion paper:                      *
C "Classical-to-critical crossovers from field  *
c theory" by C. Bagnuls and C. Bervillier       *
c************************************************
c The important subroutine is 
c fns(res,iphase,ibound,ifn,tau,theta,ikont)
c res is the return value of the subroutine
c iphase controls the phase: 
c       iphase=1 corresponds to T>Tc
c       iphase=2 corresponds to T<Tc
c ifn controls the type of function chosen:
c       ifn=1 gives the inverse correlation length for T>Tc if iphase=1
c otherwise (iphase=2) it gives the coexistence curve
c       ifn=2 gives the inverse susceptibility
c       ifn=3 gives the specific heat
c ibound controls the accuracy of the theoretical calculation:
c       ibound=1 corresponds to "max"
c       ibound=2 corresponds to "min"
c tau is |T-Tc|/Tc
c theta is the adjustable parameter which relates the scaling field t to tau:
c     (in principle t=theta*tau but for practical use we have separated the asymptotic
c      pure scaling from the correction-to-scaling contributions, see the companion paper,
c      for theta=1 one recovers the pure theoretical
c      functions of the scaling field t)
c ikont is an integer which allows to choose the nature of res
c      if ikont=0 then res gives the functions as indicated above
c      if ikont=1 then res corresponds to the effective exponent associated to the function
c          chosen according to the criteria described above. 
c All the functions are for Ising-like model (n=1)
C********************************
      implicit double precision (a-h,o-z)       
C*************************************************************
C Evolution of the functions and of the effective exponents  *
C  in terms of the pure scaling field t                      *
C  (i.e., theta=1)                                           *
C*************************************************************
C
C***
C Selection of the function
C***
      iphase=2    ! controls the phase: 1=T>Tc, 2=T<Tc
      ibound=1    ! controls the accuracy bound : 1=max, 2=min
      ifn=3       ! controls the fn: 1=xi, 2=chi, 3=c; for T<Tc, 1=coex curve
C***
C End of selection
C***
C***
C Files for saving the results
C***
      open(29,FILE='expeff.dat')
      open(30,FILE='Funct.dat')
      theta=1                    ! gives the theoretical evolution in terms of the scaling field t
      t=1.d-9
      do 30 i=1,45
      call fns(f3,iphase,ibound,ifn,t,theta,0)   !Call of the function with ikont=0
      write(30,666) t,f3
      call fns(exp2,iphase,ibound,ifn,t,theta,1)   ! Call of the effective exponent with ikont=1
      print *, t,f3,exp2
      alogt=dlog10(t)
      write(29,666) alogt,exp2     ! Ready for a Log-Lin plot
      t=t*2
30    continue
      close(29)
      close(30)
      stop   
666   format(2(G13.6,1x))
      end
C***
C Theoretical functions for n=1
C***
      SUBROUTINE fns(res,i,j,k,t,theta,ikont)
c*************************************************
c if ikont=0: res gives the function
c if ikont=1: res gives the effective exponent
c Table f contains the values of the functions' parameters.
c At the end of each set a comment allows to distinguish the
c kind of function concerned.
c For each set, the first value is the critical exponent value
c  (or minus this value for the specific heat), the second value
c  is the inverse of the critical amplitude for xi and chi, the
c  critical amplitude for aim and c, the third value is the critical
c  background which is non zero only for c.
c The two values of the subcritical exponent delta corresponding to the two 
c bounds "max" and "min" are contained in the table del
c*************************************************
      implicit double precision (a-h,o-z)
      dimension del(2),x(5),y(5)
      dimension f(15,3,2,2)
      data del/0.49125,0.50031/
      data F/0.630501,2.12411,0.,16.87409,16.72772,23.51302,-0.2154827,
     #0.6009808,-1.955281D-03,5.460931,-4.356399D-02,0.,0.,0.,0.,         !ximax  T>Tc (corr length)
     #1.24194,3.80403,0.,10.7784,10.6849,18.2831,-0.447952,0.211862,
     #-1.12883D-03,2.82005,-3.47989D-02,0.,0.,0.,0.,                      !chimax  T>Tc (suscept)
     #-.108496,1.74928,-3.87997,22077.0,21885.5,8.15287,-8.12340D-03,
     #36.4432,0.223350,395.406,1.76520D-03,0.,0.,0.,0.,                   !cmax  T>Tc (specif heat)
     #0.629121,2.09256,0.,1.793553,1.794109,32.37368,-0.1214804,
     #11.42729,-0.1263489,2.192044,-1.041270D-02,0.,0.,0.,0.,             !ximin  T>Tc (corr length)
     #1.239485,3.706359,0.,1.229767,1.230149,2.198050,-2.015454D-02,
     #24.55992,-0.2546755,10.44460,-0.2041400,0.,0.,0.,0.,                !chimin  T>Tc (suscept)
     #-.112636,1.58819,-3.57607,69.75599,69.77762,11.38361,
     #-1.004983D-01,11.37558,8.030914D-02,33.34772,0.2454611,0.,0.,0.,
     #0.,                                                                 !cmin  T>Tc (specif heat)
     #0.32516,0.9213251,0.,64.93122,64.36800,1.770551,1.028455D-02,
     #46.28114,0.1928966,89.74606,-5.501773D-02,10.80746,0.2015166,0.,
     #0.,                                                                 !aimax  T<Tc (coex curve)
     #1.24194,19.19139,0.,8.662018,8.586883,8.293155,0.2623473,
     #37.19002,-1.070686,71.02634,-2.254137,62.54640,2.578595,
     #0.,0.,                                                              !chimax  T<Tc (suscept)
     #-.108496,3.184180,-3.879732,49.56056,49.13067,4.262502,
     #6.145600D-03,66.44198,0.1076623,20.11234,1.031841D-01,0.,0.,0.,0.,  !cmax  T<Tc (specif heat)
     #0.32357,0.8971946,0.,13.29911,13.30323,131.5962,2.430934D-02,
     #1.304663,4.734644D-03,23.57031,0.1913076,8.270542,0.1325084,0.,0.,  !aimin  T<Tc (coex curve)
     #1.239485,16.78313,0.,360.4357,360.5474,22.85732,0.4525390,
     #204.6138,5.788026D-02,3.581834,6.527502D-02,41.70674,-1.054664,
     #0.,0.,                                                              !chimin  T<Tc (suscept)
     #-0.112636,3.004344,-3.576325,225.67103,225.74101,13.429043,
     #7.624208D-02,183.93336,-29.012144,56.400856,0.16278169,183.93769,
     #28.998392,0.,0./                                                    !cmin  T<Tc (specif heat)
C*********
C  definitions of constants
C*********
      d=del(j)    ! correction exponent
      e=F(1,k,j,i) ! asymptotic exponent
      z=f(2,k,j,i)  ! asymptotic amplitude 
      x6=f(3,k,j,i) ! additive critical background
      s1=f(4,k,j,i)
      s2=f(5,k,j,i)
      kmax=0
      do 1 ii=1,5
      x(ii)=f(2*ii+4,k,j,i)    ! definitions of the Xi's and Yi's
      if(x(ii).eq.0.) go to 2
      y(ii)=f(2*ii+5,k,j,i)
1     kmax=kmax+1
2     continue
      if(ikont.eq.0) then    ! calculation of the function
      res=Z*t**e
      if(theta.eq.0.) then   ! If theta=0
      res=res+x6             ! there is no correction-to-scaling
      return                 ! only the pure scaling form
      end if                 ! survives
      tt=theta*t
      D=D-1+(S1*dsqrt(tt)+1)/(S2*dsqrt(tt)+1)
      trr=(tt)**D
      do 3 kk=1,kmax
3     res=res*(1+X(kk)*trr)**Y(kk)
      res=res+X6
      return
      end if
      if(ikont.eq.1) then    ! calculation of the effective exponent
      if(theta.eq.0) then   ! If theta=0
      res=e                 ! then
      if(k.eq.3) res=-res   ! the effective exponent
      return                ! reduces to the constant 
      end if                ! critical exponent
      tt=theta*t
      D=D-1+(S1*dsqrt(tt)+1)/(S2*dsqrt(tt)+1)
      trr=(tt)**D
      Dp=(S1-S2)/(2*dsqrt(tt)*(S2*dsqrt(tt)+1)**2)
      res=0.
      do 4 kk=1,kmax
4     res=res+X(kk)*Y(kk)/(1+X(kk)*trr)
      res=res*(Dp*tt*DLOG(tt)+D)*trr
      res=res+e
      if(k.eq.3) res=-res     ! the exponent changes sign for c
      return
      end if
      print *, "Error, ikont is ",ikont, ", but must be 0 or 1."
      stop
      end
\end{verbatim}

\newpage
\section{Computer Program in Fortran \\ for the $n$-vector model ($n=1, 2, 3$) in the homogeneous phase
\\ (former criteria)}
\begin{verbatim}
      program FTcrossN
c************************************************
c            former                             *
c Crossover functions up to six loop orders     *
c************************************************
c For N-vector-like systems (n=1, 2, 3)         *
c in the homogeneous phase only                 *
c************************************************
c see the companion paper:                      *
C "Classical-to-critical crossovers from field  *
c theory" by C. Bagnuls and C. Bervillier       *
c************************************************
c The important subroutine is 
c fns(res,iN,ibound,ifn,tau,theta,ikont)
c res is the return value of the subroutine
c iN controls the value of n: 
c       iN=1 corresponds to n=1
c       iN=2 corresponds to n=2
c       iN=3 corresponds to n=3
c ifn controls the type of function chosen:
c       ifn=1 gives the inverse correlation length
c       ifn=2 gives the inverse susceptibility
c       ifn=3 gives the specific heat
c ibound controls the accuracy of the theoretical calculation:
c       ibound=1 corresponds to "max"
c       ibound=2 corresponds to "min"
c tau is |T-Tc|/Tc
c theta is the adjustable parameter which relates the scaling field t to tau:
c     (in principle t=theta*tau but for practical use we have separated the asymptotic
c      pure scaling from the correction-to-scaling contributions, see the companion paper,
c      for theta=1 one recovers the pure theoretical
c      functions of the scaling field t)
c ikont is an integer which allows to choose the nature of res
c      if ikont=0 then res gives the functions as indicated above
c      if ikont=1 then res corresponds to the effective exponent associated to the function
c          chosen according to the criteria described above. 
C********************************
      implicit double precision (a-h,o-z)       
C*************************************************************
C Evolution of the functions and of the effective exponents  *
C  in terms of the pure scaling field t                      *
C  (i.e., theta=1)                                           *
C*************************************************************
C
C***
C Selection of the function
C***
      iN=2        ! controls the value of n: 1, 2, or 3
      ibound=1    ! controls the accuracy bound : 1=max, 2=min
      ifn=3       ! controls the fn: 1=xi, 2=chi, 3=c
C***
C End of selection
C***
C***
C Files for saving the results
C***
      open(29,FILE='expeff.dat')
      open(30,FILE='Funct.dat')
      theta=1                    ! gives the theoretical evolution in terms of the scaling field t
      t=1.d-9
      do 30 i=1,45
      call fns(f3,iN,ibound,ifn,t,theta,0)   !Call of the function with ikont=0
      write(30,666) t,f3
      call fns(exp2,iN,ibound,ifn,t,theta,1)   ! Call of the effective exponent with ikont=1
      print *, t,f3,exp2
      alogt=dlog10(t)
      write(29,666) alogt,exp2     ! Ready for a Log-Lin plot
      t=t*2
30    continue
      close(29)
      close(30)
      stop   
666   format(2(G13.6,1x))
      end
C***
C Theoretical functions for n-vector-like systems
C***
      SUBROUTINE fns(res,i,j,k,t,theta,ikont)
c*************************************************
c if ikont=0: res gives the function
c if ikont=1: res gives the effective exponent
c Table f contains the values of the functions' parameters.
c At the end of each set a comment allows to distinguish the
c kind of function concerned.
c For each set, the first value is the critical exponent value
c  (or minus this value for the specific heat), the second value
c  is the inverse of the critical amplitude for xi and chi, the
c  critical amplitude for c, the third value is the critical
c  background which is non zero only for c.
c The six values of the subcritical exponent delta corresponding to the two 
c bounds "max" and "min" for each value of n are contained in the table del
c*************************************************
      implicit double precision (a-h,o-z)
      dimension del(2,3),x(5),y(5)
      dimension f(15,3,2,3)
      data del/0.49125,0.50031,0.52012,0.52737,0.54975,0.55043/    
      data F/0.630501,2.12411,0.,16.87409,16.72772,23.51302,-0.2154827,
     #0.6009808,-1.955281D-03,5.460931,-4.356399D-02,0.,0.,0.,0.,       !ximax T>Tc n=1 (corr length)
     #1.24194,3.80403,0.,10.7784,10.6849,18.2831,-0.447952,0.211862,
     #-1.12883D-03,2.82005,-3.47989D-02,0.,0.,0.,0.,                    !chimax T>Tc n=1 (suscept)
     #-.108496,1.74928,-3.87997,22077.0,21885.5,8.15287,-8.12340D-03,
     #36.4432,0.223350,395.406,1.76520D-03,0.,0.,0.,0.,                 !cmax T>Tc n=1 (specif heat)
     #0.629121,2.09256,0.,1.793553,1.794109,32.37368,-0.1214804,
     #11.42729,-0.1263489,2.192044,-1.041270D-02,0.,0.,0.,0.,           !ximin  T>Tc n=1 (corr length)
     #1.239485,3.706359,0.,1.229767,1.230149,2.198050,-2.015454D-02,
     #24.55992,-0.2546755,10.44460,-0.2041400,0.,0.,0.,0.,              !chimin  T>Tc n=1 (suscept)
     #-.112636,1.58819,-3.57607,69.75599,69.77762,11.38361,
     #-1.004983D-01,11.37558,8.030914D-02,33.34772,0.2454611,0.,0.,0.,
     #0.,                                                               !cmin  T>Tc n=1 (specif heat)
     #0.669848,2.578296,0.,28.84310,29.43534,15.75544,-0.2585115,
     #5.012764,-4.343250D-02,77.73647,-3.775196D-02,0.,0.,0.,0.,        !ximax T>Tc n=2 (corr length)
     #1.31792,5.51529,0.,63.8357,65.1464,16.9739,-0.508216,
     #1.85577,-1.44706D-02,7.33240,-1.03476D-01,247.042,-9.67699D-03,
     #0.,0.,                                                            !chimax T>Tc n=2 (suscept)
     #9.488051D-03,-85.85105,80.36121,457.8256,467.2262,21.93721,
     #-4.630859D-02,11.62941,-9.686347D-03,15.18151,3.701884D-02,
     #0.,0.,0.,0.,                                                      !cmax T>Tc n=2 (specif heat)
     #0.6678685,2.527959,0.,92.02055,94.61002,1.400964,-3.366864D-03,
     #5.520135,-4.367003D-02,395.4183,-4.673720D-03,18.17701,
     #-0.2840264,0.,0.,                                                 !ximin  T>Tc n=2 (corr length)
     #1.314001,5.307447,0.,60.60521,62.31065,15.15227,-0.5613569,
     #2.473990,-2.203622D-02,5.819209,-3.266531D-02,246.3158,
     #-1.194353D-02,0.,0.,                                              !chimin  T>Tc n=2 (suscept)
     #3.603233D-03,-207.0019,201.9794,367.4423,377.7822,17.09462,
     #-12.10980,3.267621,-1.663525D-04,17.08843,12.10276,0.,0.,0.,0.,   !cmin  T>Tc n=2 (specif heat)
     #0.7056167,3.014192,0.,76.15010,80.13691,4.745939,-8.777504D-02,
     #445.7242,-7.186912D-03,8.077017,0.2125871,13.25811,-0.5288586,
     #0.,0.,                                                            !ximax T>Tc n=3 (corr length)
     #1.389161,7.591932,0.,27.84825,29.30624,10.79849,-0.6921719,
     #5.606645D-02,5.519555D-04,316.7178,-8.449968D-03,67.73494,
     #-7.825108D-02,0.,0.,                                              !chimax T>Tc n=3 (suscept)
     #0.1168480,-19.46045,8.971331,69.91117,73.57134,389.0545,
     #-7.759979D-03,15.08428,-1.794460,0.2445648,3.624132D-04,
     #14.03141,1.568162,0.,0.,                                          !cmax T>Tc n=3 (specif heat)
     #0.7038317,2.969913,0.,78.53203,82.70273,12.80904,-0.5435931,
     #5.710801,-0.3817143,463.4073,-6.842927D-03,6.792342,0.5244869,
     #0.,0.,                                                            !ximin  T>Tc n=3 (corr length)
     #1.383616,7.231499,0.,48.19012,50.74942,12.44953,-0.7397974,
     #0.5892605,0.2315015,0.6046381,-0.2390366,260.9201,
     #-1.990006D-02,0.,0.,                                              !chimin  T>Tc n=3 (suscept)
     #0.1114902,-19.09212,9.185812,2.689863,2.832717,40.48173,
     #-0.1850101,1.173960,-0.1448486,1.743464D-02,6.113689D-04,
     #0.8190587,0.1062669,0.,0./                                        !cmin  T>Tc n=3 (suscept)
C*********
C  definitions of constants
C*********
      if(i.gt.3.or.i.lt.1) then
      print *, "Error, n= ",i, "is not available. Sorry. (0<n<4)"
      stop
      end if
      d=del(j,i)    ! correction exponent
      e=F(1,k,j,i) ! asymptotic exponent
      z=f(2,k,j,i)  ! asymptotic amplitude 
      x6=f(3,k,j,i) ! additive critical background
      s1=f(4,k,j,i)
      s2=f(5,k,j,i)
      kmax=0
      do 1 ii=1,5
      x(ii)=f(2*ii+4,k,j,i)    ! definitions of the Xi's and Yi's
      if(x(ii).eq.0.) go to 2
      y(ii)=f(2*ii+5,k,j,i)
1     kmax=kmax+1
2     continue
      if(ikont.eq.0) then    ! calculation of the function
      res=Z*t**e
      if(theta.eq.0.) then   ! If theta=0
      res=res+x6             ! there is no correction-to-scaling
      return                 ! only the pure scaling form
      end if                 ! survives
      tt=theta*t
      D=D-1+(S1*dsqrt(tt)+1)/(S2*dsqrt(tt)+1)
      trr=(tt)**D
      do 3 kk=1,kmax
3     res=res*(1+X(kk)*trr)**Y(kk)
      res=res+X6
      return
      end if
      if(ikont.eq.1) then    ! calculation of the effective exponent
      if(theta.eq.0) then   ! If theta=0
      res=e                 ! then
      if(k.eq.3) res=-res   ! the effective exponent
      return                ! reduces to the constant 
      end if                ! critical exponent
      tt=theta*t
      D=D-1+(S1*dsqrt(tt)+1)/(S2*dsqrt(tt)+1)
      trr=(tt)**D
      Dp=(S1-S2)/(2*dsqrt(tt)*(S2*dsqrt(tt)+1)**2)
      res=0.
      do 4 kk=1,kmax
4     res=res+X(kk)*Y(kk)/(1+X(kk)*trr)
      res=res*(Dp*tt*DLOG(tt)+D)*trr
      res=res+e
      if(k.eq.3) res=-res  ! the exponent changes sign for c
      return
      end if
      print *, "Error, ikont is ",ikont, ", but must be 0 or 1."
      stop
      end
\end{verbatim}


\begin{references}
\bibitem{foot18} The present calculations differ from \cite{4073,4323} in that:
1) our treatment is more accurate, 2) the specific heat is considered in the two phases,
3) three values of $n$ are treated.
\bibitem{foot1}  That is precisely how universal combinations of amplitudes
and scaling relations between exponents occur in the field theoretical
framework.
\bibitem{foot2}  Only one family of corrections to scaling (controlled by
the unique exponent $\Delta $) is accounted for.
\bibitem{foot4}  Generally speaking this is true, but $\theta $ may be nil
because there exist systems which do not approach the Ising fixed point
along the RG trajectory which links the Gaussian and Ising fixed points. For
example some models may have correction-to-scaling
terms strictly different from those accounted for by field theory. Also some
may have ``negative'' correction-to-scaling terms. See \cite{38}.
\bibitem{foot5}  The comparison is made easier than with functions having a
limited range of applicability.
\end{references}
\end{document}